\UseRawInputEncoding
\documentclass{aa} 
\usepackage{longtable}
\usepackage{booktabs}
\usepackage[utf8]{inputenc}
\usepackage{appendix}
\usepackage{graphicx}
\usepackage{subcaption} 
\usepackage{txfonts}
\usepackage{orcidlink}
\usepackage[normalem]{ulem}
\usepackage{soul}
\usepackage{cancel}
\usepackage{hyperref}
\usepackage[normalem]{ulem}
\usepackage{outlines}
\usepackage{multirow}
\usepackage{lineno}
\usepackage{amsmath} 
\usepackage{enumitem}
\usepackage{booktabs}
\usepackage{float}
\usepackage{changes}
\usepackage{soul}
\usepackage{changepage}

\usepackage[table]{xcolor} 
\newcommand{\Mjup}{\,\ensuremath{\mathrm{M}_\mathrm{Jup}}\,}

\newcommand{\Msun}{\,\ensuremath{{\rm M}_{\odot}}\,}

\definecolor{darkred}{RGB}{139,0,0}
\newcommand{\numallthousandau}{2103}
\newcommand{\alphaallthousandau}{-0.21^{+0.31}_{-0.27}}
\newcommand{\betaallthousandau}{-1.38\pm0.32}
\newcommand{\logfallthousandau}{-1.97^{+0.11}_{-0.11}}
\newcommand{\numfgkthousandau}{747}
\newcommand{\alphafgkthousandau}{0.01^{+0.48}_{-0.40}}
\newcommand{\betafgkthousandau}{-1.46\pm0.43}
\newcommand{\logffgkthousandau}{-1.71^{+0.14}_{-0.15}}
\newcommand{\nummthousandau}{1356}
\newcommand{\alphamthousandau}{-0.43^{+0.43}_{-0.36}}
\newcommand{\betamthousandau}{-1.45\pm0.51}
\newcommand{\logfmthousandau}{-2.13^{+0.16}_{-0.17}}

\newcommand{\numAUInfthousandau}{1\,000} 
\newcommand{\numAUSupthousandau}{20\,000} 
\newcommand{\numIntialthousandau}{2\,126}     
\newcommand{\numnoparallaxthousandau}{2}   
\newcommand{\numhighproperthousandau}{21} 
\newcommand{\ffractionthousandau}{1.07^{+0.31}_{-0.24}\,\%}

\begin{document}

\title{Statistics on the population of distant substellar companions to nearby stars}
 
\author{
Souleymane Ouedraogo\inst{1,2,3,*}\orcidlink{0009-0006-8817-5930},
\'Etienne Artigau\inst{3,4}\orcidlink{0000-0003-3506-5667},
Si\'e Zacharie Kam\inst{1}\orcidlink{0000-0002-8146-0177},
Orlagh L. Creevey\inst{2}\orcidlink{0000-0003-1853-6631},
Cl\'emence Fontanive\inst{5,3,4}\orcidlink{0000-0002-2428-9932},
Neil J. Cook\inst{3}\orcidlink{0000-0003-4166-4121},
Ren\'e Doyon\inst{3,4}\orcidlink{0000-0001-5485-4675},
Fabrice Bado\inst{1,2}\orcidlink{0000-0001-9696-6376}
}

\institute{
\inst{1}Laboratoire de Physique et de Chimie de l’Environnement , Universit\'e Joseph Ki-Zerbo, 03 BP 7021 Ouagadougou 03, Burkina Faso\\
\inst{2}Universit\'e C\^ote d'Azur, Observatoire de la C\^ote d'Azur, CNRS, Laboratoire Lagrange, France\\
\inst{3}Institut Trottier de recherche sur les exoplan\`etes, D\'epartement de Physique, Universit\'e de Montr\'eal, Montr\'eal, Qu\'ebec, Canada\\
\inst{4}Observatoire du Mont-M\'egantic, Qu\'ebec, Canada\\
\inst{5}SUPA, Institute for Astronomy, University of Edinburgh, Blackford Hill, Edinburgh EH9 3HJ, UK\\
\inst{*}\email{souley.ouedraogo@ujkz.bf}
}

\date{Accepted 17  July 2026 }

\titlerunning{Population of Distant Substellar Companions}
\authorrunning{Ouedraogo et al.}
 
\abstract
{Wide substellar companions to  stars provide valuable benchmarks for investigating the formation and evolutionary pathways of brown dwarfs and giant planets far beyond the extent of protoplanetary discs. Association to their primary also provides age and metallicity constraints that are otherwise challenging to obtain for isolated objects.
}
{We present a statistical approach to determine the properties of the population of substellar companions to nearby main-sequence stars. This work uses proper motion measurements derived from multi-epoch WISE imaging. Our sample includes {\numallthousandau} stars within 20\,pc and has a sensitivity sufficient to detect substellar companions down to $\sim$300\,K and out to orbital separations of \hbox{\numAUSupthousandau\,AU}.}
{ We use a Bayesian approach, based on Markov Chain Monte Carlo (MCMC) methods, to constrain the parameters of the companion mass and semi-major axis distributions. We account for observational biases, notably the decreasing detectability of fainter objects with increasing absolute magnitude M$_{W2}$, and we robustly explore the underlying population properties.
We assume that the companion mass (M) and semi-major axis (a) { follow power laws, described by} \(
{  \textit{d}^2\textit{n} \propto \textit{M}^{\alpha} \textit{a}^{\beta} \textit{dM} \,\textit{da}}
\).}
{Fitting this model to the data yields the following constraints at the 1$\sigma$ confidence level { $\alpha = \alphaallthousandau$ and
$\beta = \betaallthousandau$ and an occurrence rate of
$\ffractionthousandau$ for companions between 5 and 80\,M$_J$ and at separations of {\numAUInfthousandau--\numAUSupthousandau\,AU.}} The occurrence rate of companions increases for tighter orbits, and more massive host stars.}
{ The inferred mass properties of the substellar companion population are broadly consistent with previous results from direct imaging surveys, despite differences in target selection. {Notably, the power-law distributions in both mass and separation are consistent with those of stellar binaries}. As the input sample is about an order of magnitude larger than that of any previous imaging survey of substellar companions, we significantly improve the statistical characterisation of these objects. Our results suggest that the wide-separation companions represent the low-mass tail of the stellar companion population, rather than a direct extension of the population of close-in giant planets.}


\maketitle
\nolinenumbers

\section{Introduction}
{\ Over the past decades, direct imaging and wide-field surveys have shown that planetary and brown dwarf (BD) companions can remain gravitationally bound to their host stars at separations ranging from a few tens to several tens of thousands of astronomical units (AU). These discoveries highlight the remarkable diversity of planetary system architectures and suggest that wide-orbit companions are a natural outcome of star and planet formation processes \citep{kratter_matzner_2006,bate_2012}.

In the Solar System, this distant region corresponds to the Oort Cloud, which marks the transition between the Sun’s gravitational dominance and the influence of the Galactic environment \citep{oort_1950}. The long-term stability of such outer reservoirs provides important constraints on their formation pathways and dynamical evolution, as well as on their susceptibility to external perturbations — including close stellar encounters, Galactic tides, and the Sun’s birth cluster conditions \citep{heisler_tremaine_1986,kaib_quinn_2009,brasser_morbidelli_2013}.

Early hypotheses invoked the presence of a massive, unseen solar companion \citep[e.g.,][]{alvarez_evidence_1984a}. However, modern dynamical studies emphasise the dominant role of external perturbations. In particular, \citet{raymond_kaib_2026} demonstrated that Galactic tides, stellar flybys, and early cluster environments can reproduce many observed properties of the distant Solar System, including the orbital distribution of long-period comets.}

While wide-field infrared surveys like 2 Micron All-Sky Survey \citep[2MASS;][]{2mass} and Wide-Field Infrared Survey Explorer \citep[WISE;][]{wise} have ruled out the presence of any such massive companion, the existence of a distant planetary-mass companion, such as the proposed Planet Nine \citep{2021Brown}, remains a possibility. These hypotheses highlight the relevance of searching for faint, wide-separation companions in the outskirts of planetary systems. This concept is further supported by observations beyond the Solar System with the discovery of BDs at separations of thousands of AU from their host stars \citep[e.g., GU Psc\,b at $\sim$2000\,AU;][]{2014Naud}. Such systems demonstrate that BDs and planetary-mass companions can and do populate the outer regions of stellar systems.

Over the past decades, numerous surveys have sought to identify BD and planetary-mass companions on wide orbits. Most campaigns used adaptive optics (AO) and have focused on young stars in nearby associations within $\sim 100$\,pc, capitalising on the higher luminosity of substellar objects at early stages of their evolution. However, these young systems should also have older, cooler counterparts among the nearby field population, whose proximity compensates for their faintness. Such Gyr-old analogues in the Solar neighbourhood are best probed in the mid-infrared where their thermal emission remains detectable despite their relatively cold temperatures.

Our knowledge of these systems stems largely from a combination of targeted searches and serendipitous discoveries {  \citep[e.g.,][]{2011Kirkpatrick,2014Luhman,2020Fontanive}}. Some companions have emerged unexpectedly in searches for isolated BDs; others were uncovered through citizen science efforts like Backyard Worlds {  \citep[e.g.,][]{2017Kuchner,Faherty_backyard_2018_a, 2021Faherty,2024_Rothermich}}. However, most of these efforts are not statistically complete, and interpreting their results requires caution. While AO surveys of young moving group (YMG) members allow for reasonably accurate age estimates {  \citep[e.g.,][]{2013Malo,2018Gagne}}, {  young star‑forming regions such as the Taurus complex or the Orion region — where pre‑main sequence populations have ages robustly constrained by isochrone fitting — also offer reliable age constraints \citep{2015Bell,2018Kounkel,2019Zari}. By contrast}, the same cannot be said for field BDs whose physical parameters often rely heavily on models and external calibrators. {  This makes individual benchmark systems, where the primary star's age and metallicity are known, especially valuable for anchoring BD evolutionary models and masses.}

{ 
Despite major progress in detecting planetary-mass companions to main-sequence stars, current techniques remain sensitive to limited ranges of orbital separation and host age. Radial-velocity and transit surveys mainly probe companions within a few AU, leaving the regime from about 5 to several thousand AU relatively unexplored, especially around older stars. We address this gap with a statistically complete, volume-limited catalogue of BD companions, accounting for selection biases and observational constraints. Wide companions may form through gravitational instability in massive discs, direct collapse in turbulent clouds \citep{2016Kratter}, or core accretion \citep{1996Pollack,2004Ida,2009Mordasini}, and can later be redistributed by stellar fly-bys \citep{Goodwin2007} or Kozai-Lidov evolution in triple systems \citep{Naoz2016}.
}

{  In this study, we use archival WISE \citep{wise} data to constrain the statistical properties of wide-orbit substellar companions, down to the planetary-mass regime, at projected separations of up to \numAUSupthousandau\,AU.

Although this survey does not present the discovery of new BD companions, its primary goal is to provide a statistically robust characterisation of the BD population around main-sequence stars. In particular, the survey is designed to constrain the mass function of BD companions, testing whether it rises towards lower masses—possibly indicating a significant population of low-mass objects scattered to wide orbits—or instead declines at low masses, as expected from the turnover of the stellar binary mass function.

Furthermore, the analysis aims to quantify the distribution of companions as a function of orbital separation, including the presence of any drop-off at large separations, and to compare the observed trends with expectations from dynamical stability arguments for wide binaries evolving in the galactic disc. Finally, by examining the companion population as a function of stellar properties, this work investigates the dependence of BD companion occurrence and characteristics on host star mass, providing constraints on the dominant formation pathways of wide-orbit substellar companions.

Sect.~\ref{survey_description} describes the stellar sample and the companion detection methodology. The results of the search are presented in Sect.~\ref{survey_results}. In Sect.~\ref{survey_sensitivity}, we derive the global completeness map for the full sample, as well as separate maps for FGK and M dwarfs. These completeness estimates are incorporated into a Markov Chain Monte Carlo (MCMC) framework in Sect.~\ref{sec:mcmc} to constrain the parameters governing the companion population. We then discuss the implications of our results and compare them with predictions from planet formation models {bf in Sect.~\ref{sec:comp}}, before summarising our conclusions in Sect.~\ref{perspectives}.
} 

\section{Survey description\label{survey_description}}

{ While several previous studies have searched for brown dwarf (BD) companions around nearby stars and characterised individual systems, none provides a volume-limited survey of substellar companions from which robust statistical conclusions can be drawn. This limitation is not restricted to imaging surveys, although larger-scale and more complete surveys exist for stellar binaries \citep[e.g., ][]{2021ElBadry}.

In the present study, we perform a search on a sample that is larger by an order of magnitude than those used in previous companion searches, with sensitivity ranging from early-L dwarfs \citep{kirkpatrick_dwarfs_1999} to well into the Y-dwarf regime \citep{cushing_discovery_2011}.} 

We opt for the 20\,pc sample of all main-sequence stars, as this volume represents an optimal balance between completeness and detectability. The \textit{Gaia} survey \citep{gaia_collaboration_vizier_2022} is complete to the bottom of the main sequence within this distance \citep{2023Golovin}, while WISE \citep{wise} provides the capability to detect late-T dwarfs \citet{burgasser_2000} {  out to distances beyond 20\,pc} and Y dwarfs to $\sim$ 10\,pc \citep[e.g., WISEP J173835.52+273258.9,][]{2011Kirkpatrick,2011Cushing}.

By construction, the stars in our sample are predominantly representative of the thin-disc stellar population, as they are drawn from the local solar neighbourhood. In contrast to surveys that have focused on restricted stellar samples, such as those carried out with GPI \citep{marois_gpi_2014}, NICI\,\citep{nici}, NaCo \citep{NaCo}, and SPHERE \citep{desidera_sphere_2021}, or the studies conducted as part of the PSYM-WIDE project \citep{projet_psym} examining 95 young stars from K5 to L5 and nearby BDs or the WEIRD project \citep{PROJET_WEIRD}, which targeted 344 stars, our approach is based on a sample nearly ten times larger ($>$2000 stars). This significant expansion provides a more statistically robust perspective on the properties of BDs as companions, particularly within stellar environments similar to that of the Sun. The main drawback of this sample selection is that stellar ages are not generally known and have to be handled statistically.

\subsection{The 20\,pc main sequence sample in Gaia}

To construct our sample, we rely on the complete and homogeneous dataset of the solar neighbourhood within a 20\,pc radius, extracted from the catalogue of \citet{kirkpatrick_2024}. This catalogue includes approximately 3600 objects, compiled using Gaia EDR3 astrometric data, complemented by additional sources such as 2MASS, WISE, Pan-STARRS, and a systematic literature review.

\begin{figure}[htb!]
\includegraphics[width=1.\linewidth]{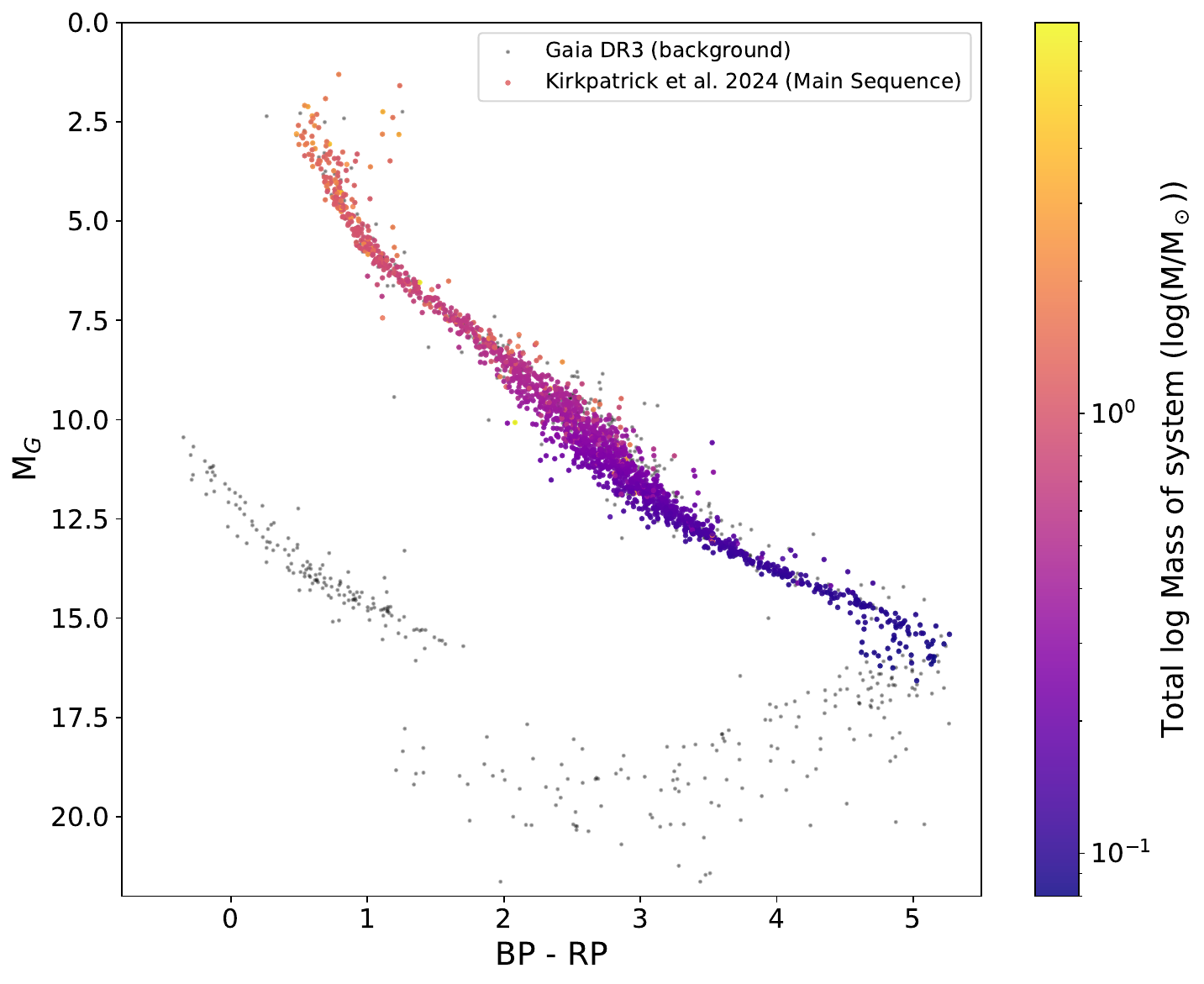}

\caption{Hertzsprung–Russell (HR) diagram of absolute magnitude $\text{M}_\text{G}$ as a function of $\text{Bp} - \text{Rp}$. The background distribution corresponds to Gaia DR3 sources located within 20\,pc of the Sun. The colour scale indicates the logarithm of the total mass of the system. The data from \citet{kirkpatrick_2024} include {~\numIntialthousandau} systems with at least one main-sequence component {  (two objects lack parallax measurements: LP 822$-$37 AB and L 128$-$7)}, which form the baseline sample used in our analysis. {We focused on stellar objects for our sample and the distinction becomes challenging at the very low-mass end without an accurate age. We applied an absolute magnitude cut at $M_G < 15.5$, corresponding to roughly spectral type M9, but acknowledge that earlier objects may be substellar and some L dwarfs are low-mass stars.}
} 
\label{fig:hr_diagramm}
\end{figure}

{Thanks to the inclusion of Gaia data, the sample is basically complete for nearby objects.} 
{  Gaia provides a precise characterisation of stellar primaries, including distances, proper motions, and absolute magnitudes. These measurements constitute a robust foundation for the identification of companions in complementary surveys, such as WISE.} The resulting catalogue serves as a benchmark for studies of the local substellar population. From this dataset, we selected {{\numIntialthousandau}} main-sequence systems (\autoref{fig:hr_diagramm}) for which astrometric, photometric, and best-estimate masses are available.

Due to the nature of our selection, our sample spans a wide range of stellar ages, rather than focusing on a specific age group. Stellar ages are estimated based on the initial mass of each system, using a statistical approach to account for their main-sequence lifetime.

\subsection{Using CatWISE as a co-moving survey}

The Backyard Worlds: Planet 9 project\footnote{\url{https://www.zooniverse.org/projects/marckuchner/backyard-worlds-planet-9}}, a citizen science initiative, has greatly contributed to the discovery of cold objects in the solar neighbourhood using data from the WISE and NEOWISE missions \citep{2024_Rothermich}.
{Notable discoveries resulting from this effort include several unusual brown dwarfs, such as WISEA J153429.75-104303.3, nicknamed “The Accident” because it was identified serendipitously in the field of another candidate rather than through a dedicated search. It was originally reported in its discovery paper \citep{Kirkpatrick2021}. Its extremely low metallicity, along with its peculiar spectroscopic features and colours, remain, to date, unique among known BDs \citep{2025Faherty}}.

The analysis was performed using data obtained with the Wide-field Infrared Survey Explorer (WISE; \citealt{wise}). WISE is a NASA Earth-orbiting mission that surveyed the entire sky simultaneously at wavelengths of 3.4, 4.6, 12, and 22~$\mu$m, referred to as the $W1$, $W2$, $W3$, and $W4$ bands, respectively. The $W1$ and $W2$ bands are of particular interest as they probe the deep methane (CH$_4$) absorption feature at 3.3~$\mu$m observed in BDs, as well as the relatively transparent region near $\sim$4.6~$\mu$m. The $W1-W2$ red colour of cold BDs, which is almost unique among astronomical sources, greatly facilitates their identification. Some of these objects are part of multiple systems {  (e.g., WISE J033605.05–014350.4 in \citealp{2023Calissendorff}; CWISEP J193518.59-154620.3 \citealp{2025DeFurio})} that include substellar or planetary-mass companions revealed through various observational techniques.
Our analysis is based on the CatWISE2020 catalogue \citep{catwise}, which provides a typical astrometric precision of $\sim$ 20\,mas/yr for sources with magnitude $W1 \approx$ 15, and $\sim$\,100\,mas/yr for $W1$\,$\approx$ 17. This level of sensitivity allows efficient detection of high proper motion objects. Taking into account the dispersion of transverse velocities in the thin-disc (generally between 20 and 30\,km/s), objects located within 20\,pc typically exhibit proper motions of 200 to 300\,mas/yr. CatWISE2020 is thus well suited to identifying a large fraction of these nearby objects. This precision is sufficient to allow for the confirmation of common proper motion for nearby wide binaries. {  The CatWISE2020 catalogue was built in an entirely automated fashion from combined \textit{WISE} and \textit{NEOWISE} datasets. Each source was detected and tracked across multiple observing epochs using an automated positional matching and astrometric fitting pipeline, as described in \citet{catwise}. This procedure yielded uniform photometric and astrometric measurements, including proper motions, for more than 1.9 billion sources.} In the present work we focus on systems out to projected separations of about \numAUSupthousandau\,AU. This upper limit reflects the approximate maximum separation at which stellar systems remain gravitationally bound over Galactic timescales, depending on their mass and the dynamical conditions of their local environment {  \citep{Weinberg}}.

\subsubsection{Cuts in colour space}

Objects within 20\,pc have very accurate parallaxes (Gaia DR3, \citealp{gaia_collaboration_vizier_2022}).
 {  For objects not included in Gaia, reliable parallaxes have been compiled by \citet{kirkpatrick_2024}}.
For this work, we assume that distance uncertainties are negligible compared to photometric uncertainties in the CatWISE catalogue for determining $M_{W2}$.

In order to identify new BD candidates, we characterise their location in the colour space defined by the $W1 $ and $W2$ filters of WISE. Cold BDs of spectral types L, T, and Y are distinguished by very red $W1-W2$ colours, typically greater than 2.0\,mag \citep{2011Kirkpatrick}. A preliminary selection of candidates was carried out using CatWISE colour information \citep{catwise}. For any given host star from the input sample, we use the Gaia parallax to transform $W2$ magnitudes into $M_{W2}$ for putative companions at the distance of the star being considered. We then compare all field objects to the $M_{W2}$ vs $W1-W2$ sequence of known LTY dwarfs and select objects that have properties that would be consistent with being a BD.
\begin{figure}[!htb]
\centering
\includegraphics[width=1.\linewidth]{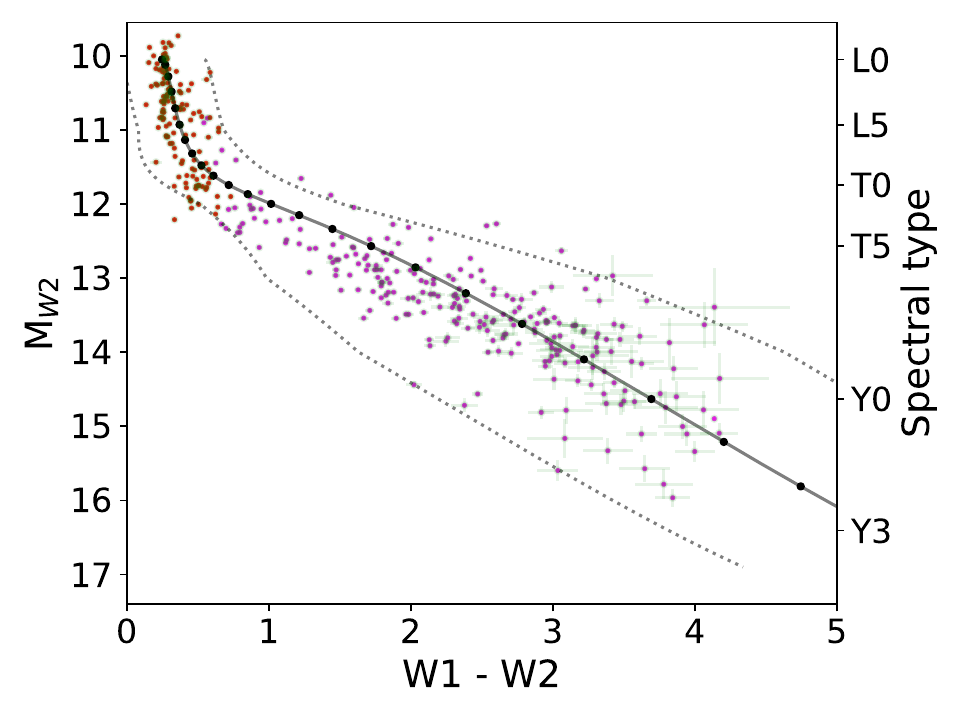}
\caption{
Colour–magnitude diagram (CMD) M$_{W2}$ vs. $W1 - W2$, including the injection of the observed BD population from \citet{kirkpatrick_2024}. The polynomial relation for BD and selection bounds for the identification of objects consistent with a companion are shown. Note that the width of the selection box increases in the late-T and Y dwarf regime as objects display an increasingly broad dispersion in their $W1-W2$ colour for a given $\rm{M}_{\rm W2}$.
} 
\label{fig:poly_diagramm}
\end{figure}

This is illustrated by the selection envelope shown in Figure~\ref{fig:poly_diagramm}, where the polynomial relation is taken from type "FLD" for M6--T9 (see Table~19 \citealp{Faherty_2016}) and type T6--Y4 (see Table~8 \citealp{2019Kirkpatrick}). 
The acceptance width for candidate selection is defined empirically,{ with a difference of $0.3$\,mag for M$_{\rm W2}<11.5$, 1.5\,mag for M$_{\rm W2}>14$ and increasing linearly between these values.} The acceptance width is defined as 4 times the median absolute deviation to the polynomial relations. Only a few field objects are retained within this selection ribbon (see left Fig.~\ref{fig:L 34-24}). Finally, we apply an absolute magnitude cut at $M_{W2} \geq 9$, which produces a realistic detection threshold dependent on spectral type: earlier-type objects must be intrinsically brighter to enter the sample, whereas later types can be detected at fainter apparent magnitudes, consistent with the survey depth of CatWISE.

\begin{figure}[htb!]
 \centering
\includegraphics[width=.9\linewidth]{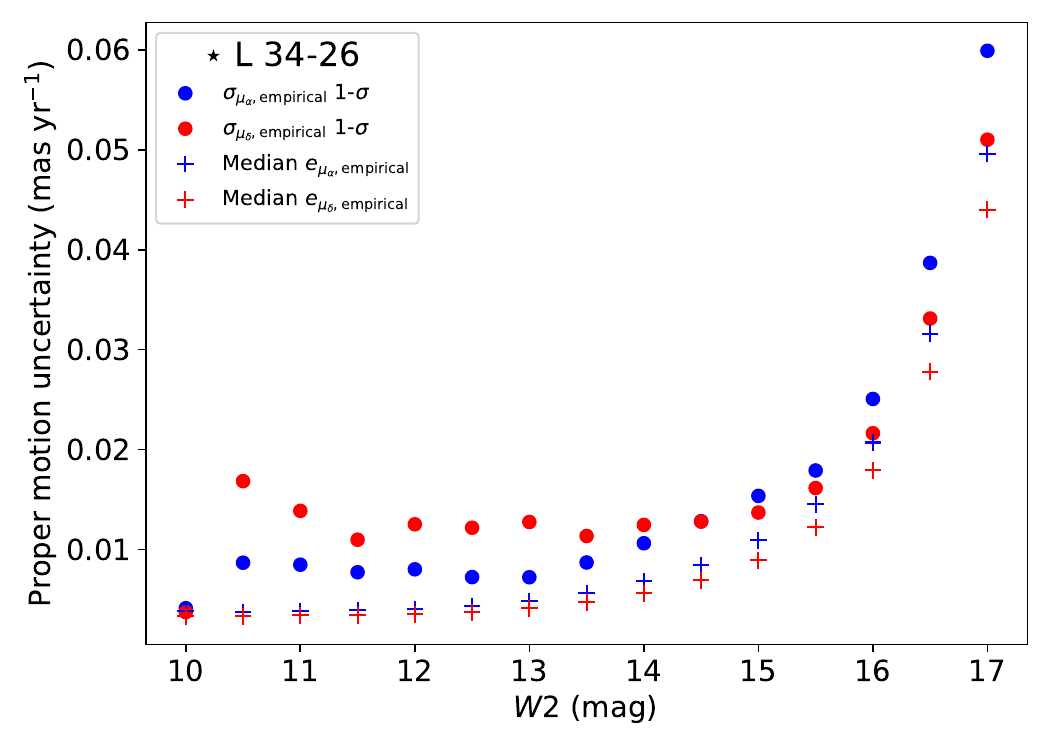}
\caption{Proper-motion dispersion as a function of $W2$ magnitude. Blue and red circles show the 1$\sigma$ uncertainties in $\mu_{\alpha,\rm empirical}$ and $\mu_{\delta,\rm empirical}$, respectively, as a function of the $W2$ bin centres for the star L~34-26. Blue and red plus signs indicate the median values of the reported uncertainties ($e_{\mu_{\alpha,\rm empirical}}$ and $e_{\mu_{\delta,\rm empirical}}$).}
\label{fig:pm_err_w2}
\end{figure}

Proper motions were extracted from the CatWISE catalogue. However, the adopted dispersion was not derived directly from the quoted individual uncertainties, but rather estimated empirically from the population of objects with comparable magnitudes, yielding a representative ($\sigma_{\mu_{\alpha},\rm empirical}$, $\sigma_{\mu_{\delta},\rm empirical}$) per spatial dimension in the sky plane. This value was derived from the $16^{\rm th}$ and $84^{\rm th}$ percentiles of the proper motion per spatial dimension and per magnitude bin (see Fig~\ref{fig:pm_err_w2}). This approach accounts not only for instrumental errors but also for the intrinsic kinematics of the sources. The measured proper motion dispersion, therefore, reflects the convolution of measurement uncertainties with the genuine motion of stellar populations, which may introduce a non-negligible level of contamination into our detection criteria.

To assess the probability that a colour candidate is part of a co-moving pair rather than a chance alignment, we analysed the regions around the host star by modeling proper motions using Gaussian distributions defined by equations \ref{G1} and \ref{G2}. To confirm that the candidate is gravitationally bound to the host star rather than a background object, we verified that its proper motion is consistent with that of the primary star {  \citep{2024_Rothermich,2021Faherty}}.

The relative {$\rho$} of the field and co-moving hypotheses are defined by
\begin{align} \rho_{\rm field} &= e^{ -0.5 \sigma^2_{\text{field}} } \label{G1}, \\ 
\rho_{\rm co-move} &= e^{-0.5 \sigma^2_{\text{co-move}} } \label{G2}, 
\end{align}
where $\sigma_{\text{field}}$ represents the observed dispersion of proper motions for objects within the field of view, in a given magnitude bin. 

The term $\sigma_{\text{co-move}}$ refers to the dispersion of the difference between proper motions reported by CatWISE and those from Gaia for the system being considered.
The corresponding quantities are defined by 
\begin{align*}
\sigma^2_{\rm field} &=\left(\frac{\mu_{\alpha,\text{wise}}}{\sigma_{\mu_{\alpha,\text{empirical}}}}\right)^2 + \left(\frac{\mu_{\delta,\text{wise}}}{\sigma_{\mu_{\delta,\text{empirical}}}}\right)^2, \\\\
\sigma^2_{\text{co-move}} &= \left(\frac{\mu_{\alpha,\text{wise}} - \mu_{\alpha\star}}{\sigma_{\mu_{\alpha,\text{wise}}}}\right)^2 + \left(\frac{\mu_{\delta,\text{wise}} - \mu_{\delta\star}}{\sigma_{\mu_{\delta,\text{wise}}}}\right)^2.
\end{align*}
The equation \ref{G1} represents the `zero-motion' or `stationary' Gaussian, characterising the proper motions of objects consistent with zero proper motion in the field frame. The equation \ref{G2}, on the other hand, represents a `co-moving' Gaussian, modeling objects gravitationally bound to their host stars. We assume that the proper motion errors in Gaia are negligible compared to those of CatWISE.

The probability that a system is truly co-moving is then estimated using the following formula: \begin{equation} P_{\text{comove}} = \frac{\rho_{\text{co-move}}\, \text{N}_2}{\rho_{\rm field}\, \text{N}_1 + \rho_{\text{co-move}}\, \text{N}_2}, \end{equation}

where $\text{N}_1$ represents the number of field objects that have colour/magnitude properties consistent with being BD candidates (e.g., objects being in the `acceptance' ribbon in Figure~\ref{fig:poly_diagramm} at the distance of the host star), and $\text{N}_2$ corresponds to the number of randomly generated and prioritised co-moving objects. To determine $\text{N}_1$, we determine the number of colour candidates we would have in a 2$^\circ$ radius around the system and scale that number to the on-sky area of a \numAUSupthousandau\,AU disc centred on the system. $\text{N}_1$ varies from near unity (more distant systems, high galactic latitudes) to a few tens. In contrast, determining $\text{N}_2$ is less straightforward; the expected number of BD companions per system is significantly below one. 

Assuming that all companions are known and dividing by the number of systems gives an $\text{N}_2\sim0.01$. We note that this may appear as a circular argument; the number of companions has an impact on the likelihood of recovering more companions. In practice, changing $\text{N}_2$ even by a factor of 2 (i.e., only half of the companions at the relevant separation would be known) does not affect our results.{  The threshold for a candidate to be considered co-moving was set at 0.5. The probability for each brown dwarf recovered in this work is given in Table~\ref{tab:stellar_parameters} in Appendix~ \ref{appendix:table}.}

\subsubsection{Construction of joint recovery map}
\label{const_map}
For all targets in the sample, we derive a recovery map in projected separation versus M$_{W2}$ space. The recovery rate as a function of projected separation is constructed from $\rho(r)$ (Eq.~\ref {eq5}) projected at the distance of the system using the parallax provided by Gaia. This determines if the object would be recovered as a source in CatWISE, but does not inform as to whether the common proper motion would be detected. In an extreme case, the host star would have a projected velocity smaller than the 1$\sigma$ limit of CatWISE at the magnitude considered.

To quantify our sensitivity to faint companions, we determined the detection limit beyond which a statistically significant measurement of common proper motion (i.e., exceeding $>$3$\sigma$) can no longer be achieved. This threshold defines the minimum brightness required for a candidate to be reliably identified as co-moving with its host star. At fainter magnitudes, the increasing uncertainties in proper motion measurements preclude a robust confirmation of the physical association.

{  In order to perform the statistical analysis presented in Sect.~\ref{sec:mcmc}, the joint map of absolute $W2$ magnitude versus projected separation must be transformed into a mass--separation map.

Since individual stellar ages are not available for our sample, and given the difficulty of determining them for a large population, we adopted a statistical approach based on the empirical method introduced by \citet{Marocco2024}. This method (see their Sect.~7.4) uses Galactic kinematics as an age proxy, while accounting for the dynamical heating of the Galactic disc observed in the solar neighbourhood \citep{Wielen1977,AumerBinney2009,Sharma2021}.

We used four independent samples of nearby stars \citep{Casagrande2011,Luck2017,Luck2018,Bensby2014}, restricted to the physical age range $0{-}13.8$~Gyr. The corresponding age probability distribution functions (PDFs) were combined to construct an age PDF representative of each star in the sample.

We assumed a simultaneous formation of the components within systems. The maximum allowable age is constrained by the main-sequence lifetime of the most massive star, estimated using the MIST stellar evolution models\footnote{\url{https://waps.cfa.harvard.edu/MIST/}} ($\log(\mathrm{Age/yr}) = 5{-}10.3$, $M = 0.1{-}300~\Msun$; \citealt{Choi_2016}, see their Fig.~13), and capped at $\sim10$~Gyr in agreement with observational constraints on the Galactic disc \citep{2009_Aumer,Haywood_2013}.

ATMO2020 \citep{atmo_2020} atmospheric and evolutionary models\footnote{\url{https://www.erc-atmo.eu/?page_id=322}}\footnote{\url{https://perso.ens-lyon.fr/isabelle.baraffe/ATMO2020/}}
 were used to interpolate companion masses from absolute $W2$ magnitudes when marginalising over the age probability distribution.

 \begin{figure*}[htb!]
 \includegraphics[width=1.\linewidth]{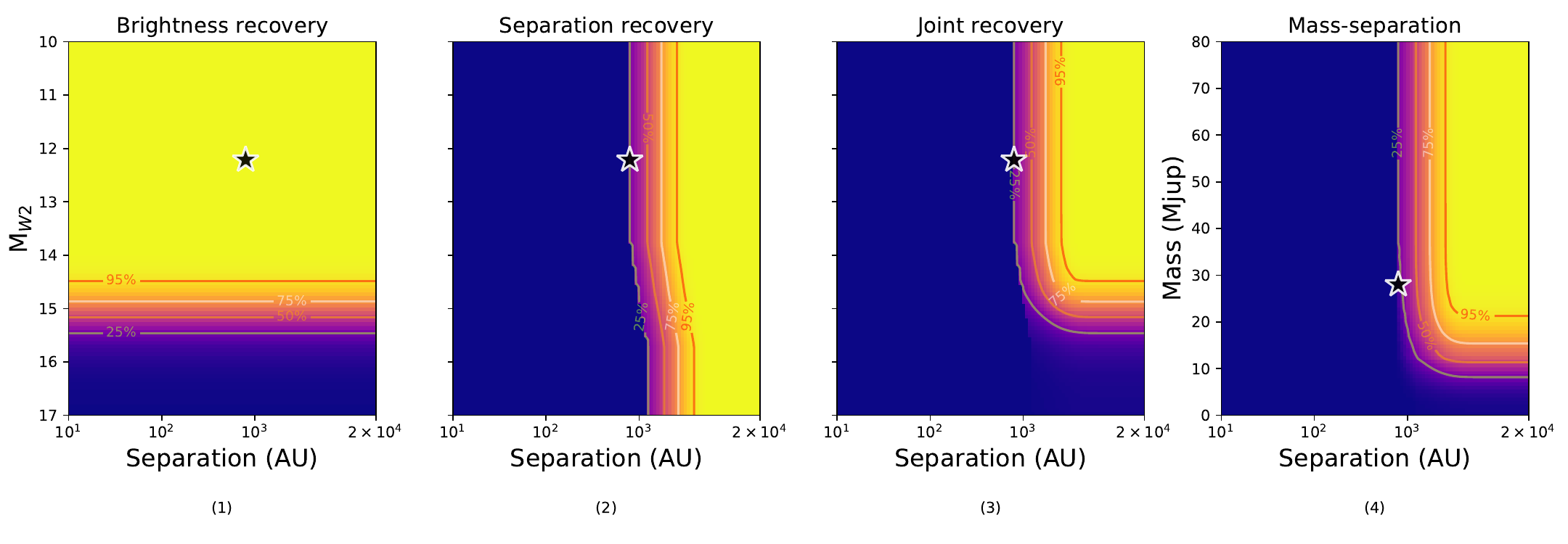}
\caption{ Completeness maps for the HN Peg system. [1]: In this recovery map, we only account for the combination of companion brightness and our ability to reliably measure its common proper motion with the host star. For the HN\,Peg system, the companion would readily be recovered and recognised as co-moving. [2] This recovery map shows reliability in the recovery of the companion at close separation to the primary, regardless of the common proper motion. In this map, the companion to HN\,Peg would have a low likelihood of recovery. [3] Joint recovery map for the system in the M$_{W2}$ vs separation diagram. [4] Same map as previous, but expressed in mass rather than $W2$ magnitude. The mass accounts for the age distribution assumed for the system (see Section~\ref{const_map}).
In all panels, the colour scale indicates the detection probability, with contours tracing levels from 10\% to 95\%. Companion injection tests show that the recovery probability is below  25\% in both absolute magnitude and mass.
\label{fig:casehnpeg}}
\end{figure*}

\begin{figure*}[!htb]
\centering

\includegraphics[width=0.9\textwidth]{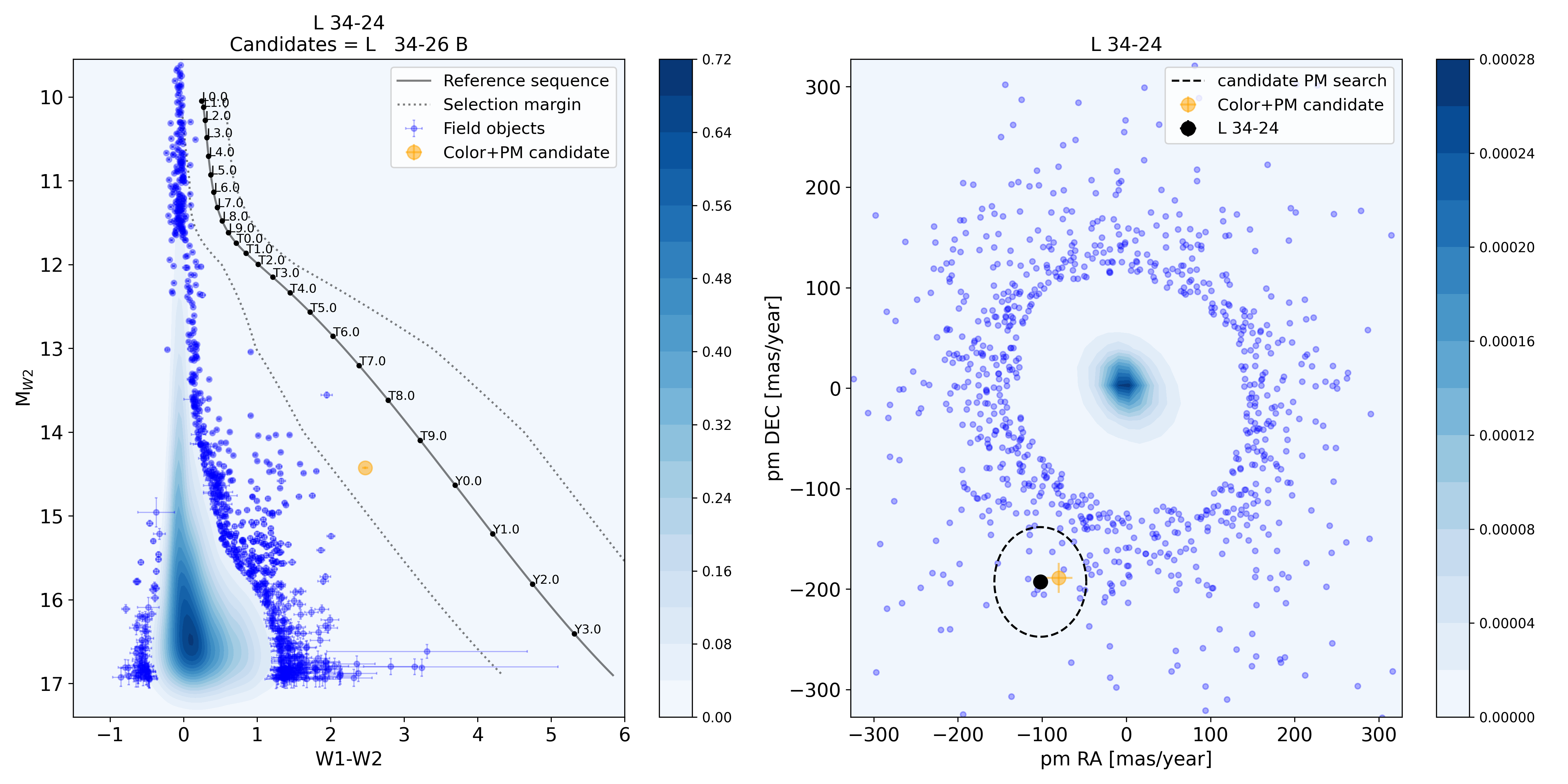}
 
\caption{Left-- CMD of the objects observed in the field of L\,34-24. Objects located inside the selection band in the left panel are considered colour candidates; typically, only a handful of such objects are found per field. In this specific case, colour candidates include the known BD companion, L\,34-24\,B, highlighted in orange. Other colour candidates include one object that would have properties of a late-T at the distance of the L\,34-24 system and a few brighter objects in part of the CMD where the BD colour sequence is close to $W1-W2 = 0$. Right-- Proper motion distribution of the objects around L\,34-24 (in black). A handful of field objects would be in the acceptance radius for co-moving objects (note that the radius is indicative here, as it is magnitude-dependent). A single object, L\,34-24\,B, is both a co-moving and a colour candidate. The other colour candidates were found to be unrelated and have a proper motion consistent with random background sources. In both panels, we show the bulk of the distribution with a colour scale, while individual high-sigma outliers are highlighted with blue points.\label{fig:L 34-24}}
\end{figure*}

Figure~\ref{fig:casehnpeg} illustrates the magnitude-limited map, the separation-limited map, and the joint recovery probability maps in the absolute $W2$ magnitude--separation and mass--separation planes for the HN\,Peg system. The T2 companion HN\,Peg\,b would be readily detectable if it were not for its angular separation from the host star intrinsic brightness and proper motion. However, given its angular separation of $43.2^{\prime\prime}$, the probability of recovering the companion drops below 25\%. Consistently, this source is not recovered in the CatWISE catalogue, in agreement with the joint detection probability inferred from our analysis.

}

\section{Survey results\label{survey_results}}

As part of our substellar companion search strategy, we systematically excluded objects with proper motions greater than \hbox{3$^{\prime\prime}$/yr} to remain within the sensitivity limits of the CatWISE2020 catalogue. For such high proper motion objects ({\numhighproperthousandau~objects, including G~272-61~AB, 61~Cyg~ AB, Barnard's Star and Epsilon Indi)}, the likelihood of recovering bound companions is significantly diminished. This selection criterion affects only a negligible fraction of the stellar population analysed. A notable example of an excluded system is the well-known triple system $\epsilon$ Indi, which consists of the primary star $\epsilon$\,Indi (HD\,209100), a K5V located at \hbox{3.64 $\pm$ 0.096\,pc}\citep{gaia_collaboration_vizier_2022}, with a high proper motion of approximately $4.7^{\prime\prime}$/yr. It is accompanied by a BD binary, $\epsilon$ Indi\,Bab (T1.5+T6), separated by $0.732^{\prime\prime}$, corresponding to a projected separation of about 2.65 AU \citep{epsilon_2004}. The binary shares a common proper motion with the host star and is widely separated at 1\,459\,AU \citep{epsilon_2003}. Although the separation of the companion system is compatible with our selection criteria, $\epsilon$ Indi represents an extreme case. Its exclusion does not significantly affect the overall mass or separation distributions in our sample.

\begin{table}[htbp]
\caption{Summary of the sample selection.}
\label{tab:sample_selection}
\centering
\begin{tabular}{lc}
\hline\hline
Selection & Number \\
\hline
Initial sample & \numIntialthousandau \\
{No parallax measurement}$^{^{(a)}}$ & {$\numnoparallaxthousandau$} \\
{High proper-motion objects}$^{{(b)}}$ & {$\numhighproperthousandau$} \\
\hline
{Final sample} & {\numallthousandau} \\
\hline
\end{tabular}
\tablefoot{
\tablefoottext{a}{Objects without available parallax measurements (see caption Fig.~\ref{fig:hr_diagramm}).}
\tablefoottext{b}{Objects with high proper motions not reliably detected in CatWISE, as described above.}
}
\end{table}

Applying
{the}
selection criteria
{listed}
in Table~\ref{tab:sample_selection},
we identified {41} objects 
{ satisfying all the conditions described in Sect.~\ref{survey_description}.}

\begin{figure*}[htb!]
\includegraphics[width=\textwidth]{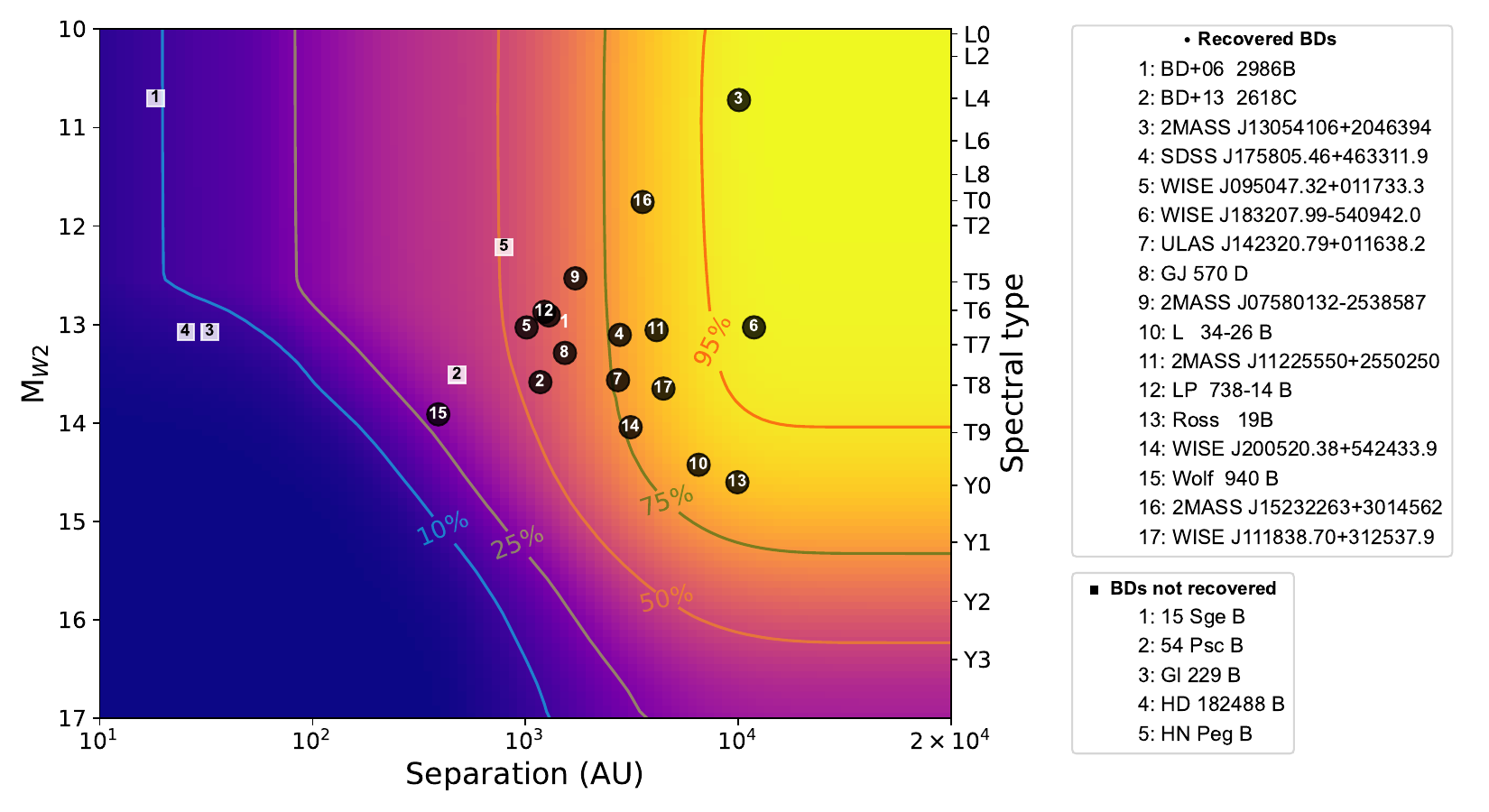}
\caption{Completeness map and average probability of detection for Gaia stars within a 20\,pc radius. It represents the average probability of detection in percent as a function of $M_{W2}$, and projected separation ($a_p$). Spectral types correspond to the reference types defined by the polynomials in Fig.~\ref{fig:poly_diagramm}. Unrecovered objects at small projected separations are not included in the figure. SCR\,J1845$-$6357~B and WISE~J072003.20$-$084651.2~B have projected separations of 4.5\,AU and 2.173\,AU, respectively.
\label{fig:carte}}
\end{figure*}

Each candidate was 
{subsequently subjected to a systematic visual inspection using all available multi-epoch unWISE/NEOWISE images displayed}
in SAOImage DS9\footnote{\url{https://sites.google.com/cfa.harvard.edu/saoimageds9}}.
{The image cubes generated from the multi-epoch observations were examined to assess the consistency of the source appearance over time, whilst catalogue information (source positions and W1/W2 photometry) was used solely to facilitate source identification. The purpose of this inspection was to verify that each candidate appeared as an isolated point source consistent with a genuine astrophysical object, and to reject imaging artefacts, extended sources, and objects affected by source blending or contamination}
{(see Appendix~\ref{appendix:false-positives} for representative examples).}

{Several candidates also exhibited optical counterparts on the blue photographic plates}
\citep{2008Lasker},
{inconsistent with the expected properties of brown dwarfs.}

{Following this validation procedure, only}
17 candidates were
{retained as credible brown dwarf candidates after comparison with archival data and the literature.}

We present in Fig.~\ref{fig:L 34-24} an example detection of the companion L\,34-24\,B. This object, initially identified as a common–proper‐motion companion to L~34-24 by \citet{2021Zhangz}, is described in detail in Appendix~\ref{appendix:recovered}, where we also provide a consistent summary of all companions recovered by our method.

Although all detected BD candidates were selected using the same criteria in our study, their physical properties were not always derived in a homogeneous manner in the original discovery papers. This heterogeneity limits the possibility of directly inferring a mass function. In contrast, the $W2$ magnitudes remain reliable and directly comparable, as they all originate from the same photometric catalogue.

The key physical parameters used in our analysis for the known and recovered systems are compiled in Appendix~ \ref{appendix:table}, Table~\ref{tab:stellar_parameters}. These include spectral types, primary masses, companion masses, parallaxes, projected separations, and WISE $W1$ and $W2$ magnitudes.

Appendix~\ref{appendix:not recovered} also presents the objects that were not recovered by our method and discusses the underlying causes.

\section{Completeness Maps and Survey sensitivity\label{survey_sensitivity}}
We used detection limits in terms of absolute magnitudes and projected separations, as illustrated in Figure~\ref{fig:casehnpeg}, to evaluate the survey's sensitivity to companions of a given magnitude and projected separations.
A simulation was first performed to produce a completeness map for each star (Fig.~\ref{fig:casehnpeg}). Summing these maps over all targets provides an estimate of the survey’s average sensitivity (Fig.~\ref{fig:carte}), expressed as the fraction of stars for which a companion of a given absolute magnitude $M_{W2}$ and projected separation would have been detected.

Figure~\ref{fig:carte} shows that the survey is most sensitive at wide separations, up to about $20{\,}000$\,AU, and down to an absolute magnitude of $M_{W2} \simeq 17$, consistent with the survey's expected sensitivity \citep{catwise}. 

We injected simulated companions and tracked those that were recovered as well as those that were not recovered (see Appendix ~\ref{appendix:not recovered}). The average detection probability reaches about 25\% for separations beyond 100\,AU and remains non-negligible (around 10\%) below this threshold. The low recovery rate at small separations results from the fact that such recovery is possible only for a handful of very nearby systems (say, within 5\,pc), which represent only a small fraction of the entire sample.

\begin{figure*}[htb!]
 \centering
 \begin{subfigure}{0.47\textwidth}
  \includegraphics[width=\linewidth]{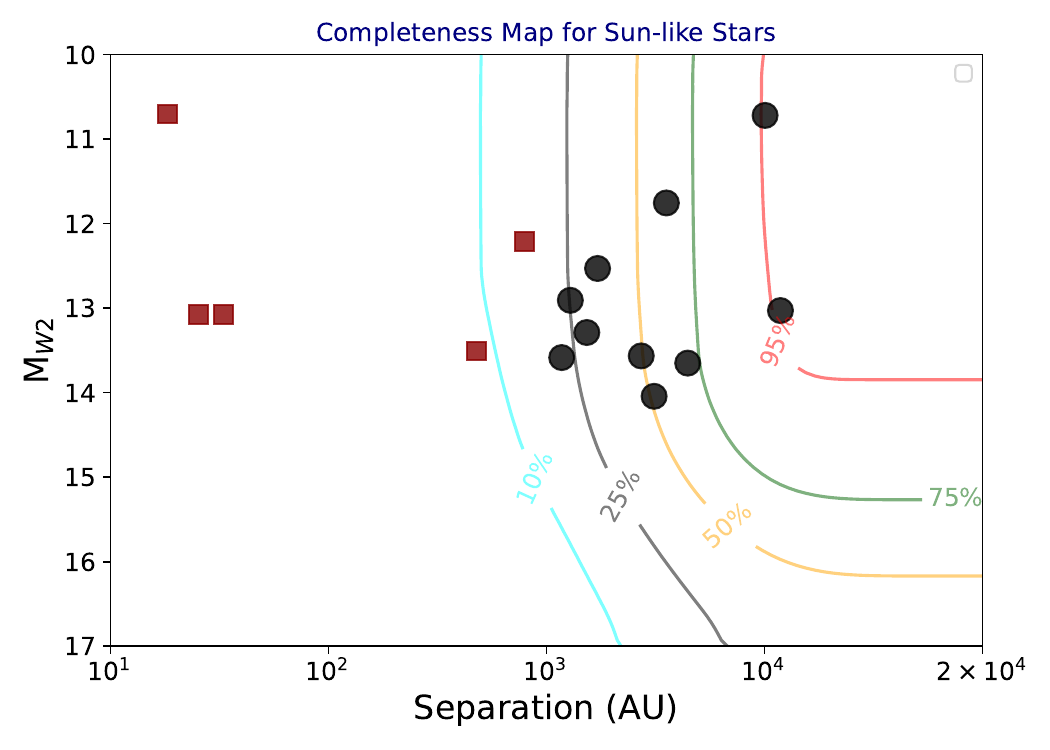}
  \caption{}
  \label{fig:FGK}
 \end{subfigure}
 \hfill
 \begin{subfigure}{0.47\textwidth}
  \includegraphics[width=\linewidth]{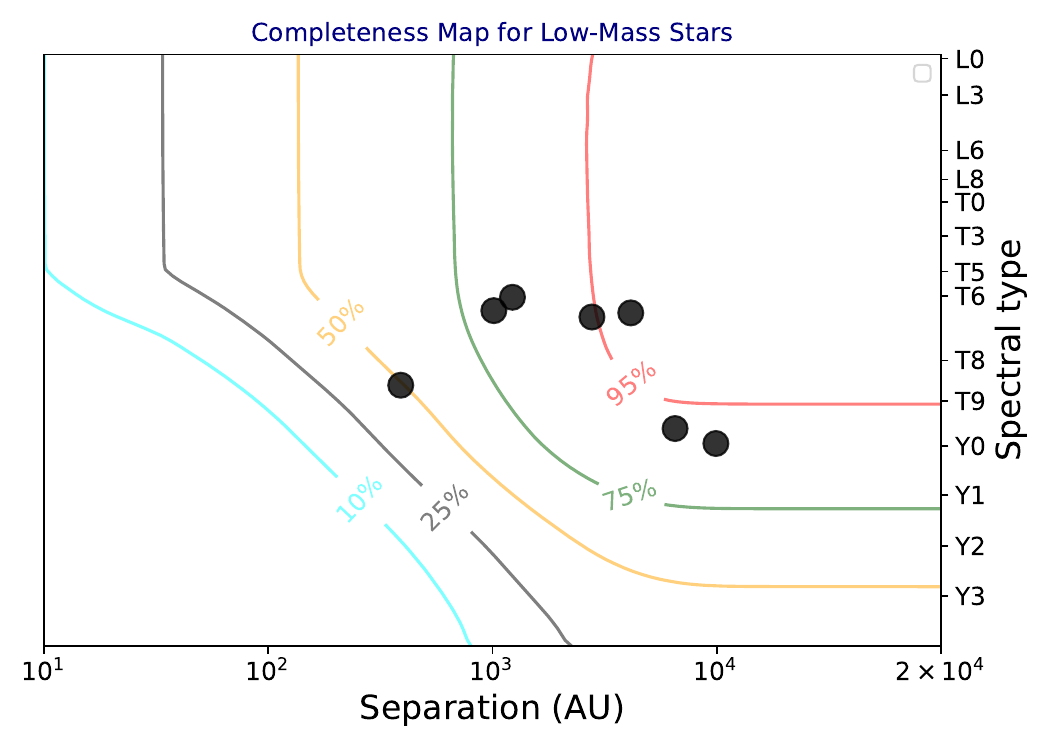}
  \caption{}
  \label{fig:M}
 \end{subfigure}

 \caption{ Completeness Map — (a) Systems with an initial total mass greater than 0.5\,\Msun (FGK stars); (b) Systems with an initial total mass lower than 0.5\,\Msun (M dwarfs). Objects marked with a square (\textcolor{darkred}{$\blacksquare$}) correspond to companions that were not recovered during our search. This map thus confirms the low detection probability for most of these objects. Those marked with a circle ($\bullet$) are the companions we successfully recovered. }
 \label{fig:prob_comp_FGKM}
\end{figure*}

\subsection{Statistical properties of Sun-like versus M dwarf companions}

In order to provide an initial assessment of how the occurrence of companions might depend on the total system mass, we divided our sample into two mass bins: above and below 0.5\,\Msun. For single-star systems, this threshold largely corresponds to the distinction between M dwarfs (lower mass) and FGK or Sun-like stars (higher mass).

{ Among the 747 FGK-type stars in our sample, we detected ten companions, corresponding to a raw detection frequency of approximately 1.3\,\%. According to the completeness map (Fig.~~\ref{fig:prob_comp_FGKM}, left panel), the detection probabilities associated with these companions mostly range between 25\% and 75\%, with no additional undetected sources found during our survey.
In contrast, seven companions were detected among the 1356 M dwarfs, corresponding to a detection frequency of 0.5\,\%. For these systems, the detection probabilities are generally greater than or equal to 50\% (Fig.~~\ref{fig:prob_comp_FGKM}, right panel).
The companion masses, compiled from the literature, range from 12.5 to 74.5\,$M_{\mathrm{Jup}}$ for FGK hosts and from 8.0 to 72.0\,$M_{\mathrm{Jup}}$ for M dwarfs.
The projected separations, computed from \textit{Gaia} parallaxes, also differ between the two subsamples: 1175–11\,853~AU for FGK stars and 390–9902~AU for M dwarfs. These differences reflect both observational sensitivity effects and the intrinsic physical properties of the systems.

{
Although the raw detection frequencies differ between the two stellar populations, the assessment on differences on intrinsic underlying distribution differences needs to account for differences in completeness between the two samples; this is done in Sect.~\ref{sec:mcmc}.
}

 \subsection{Half-life survival of companions}
 \label{sec:half-life}
We assessed the long-term dynamical stability of the identified systems. To do this, we estimated the dynamical half-life ($t_{1/2}$) of each system (see Fig.~\ref{fig:life}). This estimate is based on the total initial mass of the systems, as reported by \citet{kirkpatrick_2024}, and on their projected separations.
The calculation relies on the theoretical framework for dynamical stability in multiple systems developed by \citet{Weinberg}, which allows one to estimate the characteristic timescale beyond which a binary system may be disrupted by external gravitational interactions—such as encounters with passing stars or the Galactic tidal field. This approach enables us to evaluate the likelihood that these companions remain gravitationally bound over cosmological timescales, thereby providing insight into their long-term viability as binary or multiple systems.

 \begin{figure}[htb!]
\centering
\includegraphics[width=1.\linewidth]{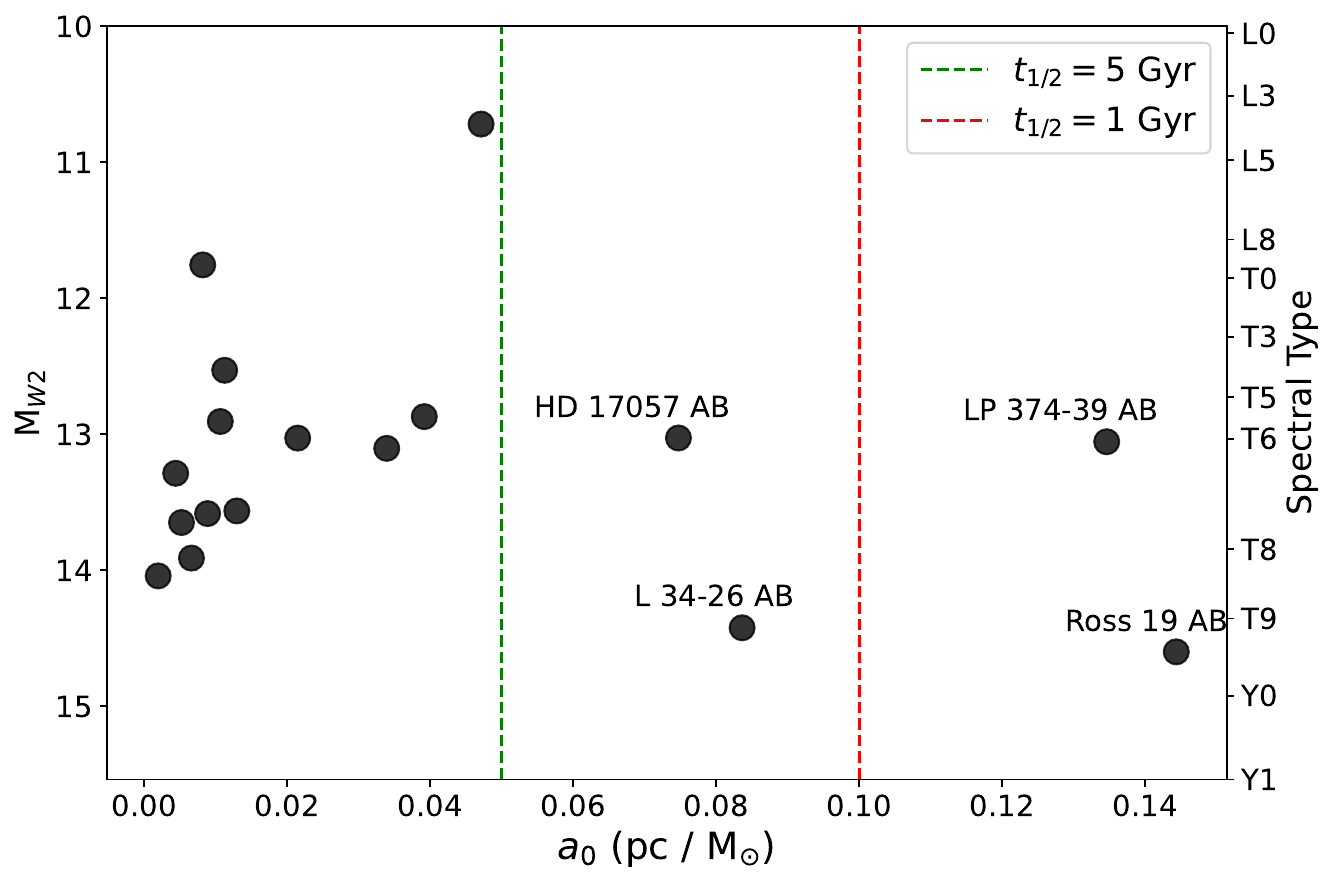}
\caption{Separation-to-total-mass ratio for L, T, and Y binaries within 20\,pc. Values above $\sim$0.05\,pc/\Msun are predicted to have a half-life of $\sim$5\,Gyr, while systems with a ratio of 0.1\,pc/\Msun have a $\sim$1\,Gyr half-life \citep{Weinberg}. We assume that the a$_0$ value scales with the separation-to-mass ratio, but note that this approximation is only strictly valid for passing stars and not giant molecular clouds (see Section IV in \citet{Weinberg} on binary mass and scaling and Figure~2 therein).}
\label{fig:life}
\end{figure}
These results reveal that four systems exhibit relatively short half-lives compared to the age of the galactic disc; two systems have half-lives of $1-5$\,Gyr and two \hbox{$<$1\,Gyr}.\\
The Sun-like system \hbox{HD 170573}, with its companion WISEJ183207.99-540942.0, has an estimated age of $>9$\,Gyr \citep{2024Calamari}, making it an ancient system. The discrepancy between its age and half-life suggests that the companion may eventually be ejected due to external perturbations, such as encounters with massive giant molecular clouds or disc stars. In its current configuration, the system would have likely been disrupted many times, given the current age of the system. This suggests that the system has been dynamically excited to its current configuration relatively recently and that it is on the cusp of disruption in the relatively near future. Conversely, the low-mass system COCONUTS-2, with companion L34-26\,B, has an estimated age of 150–800~Myr \citep{2025Zhang}, suggesting that it may have formed in its current orbital configuration.

\begin{figure*}[htb!]
\centering
\includegraphics[width=1.\linewidth]{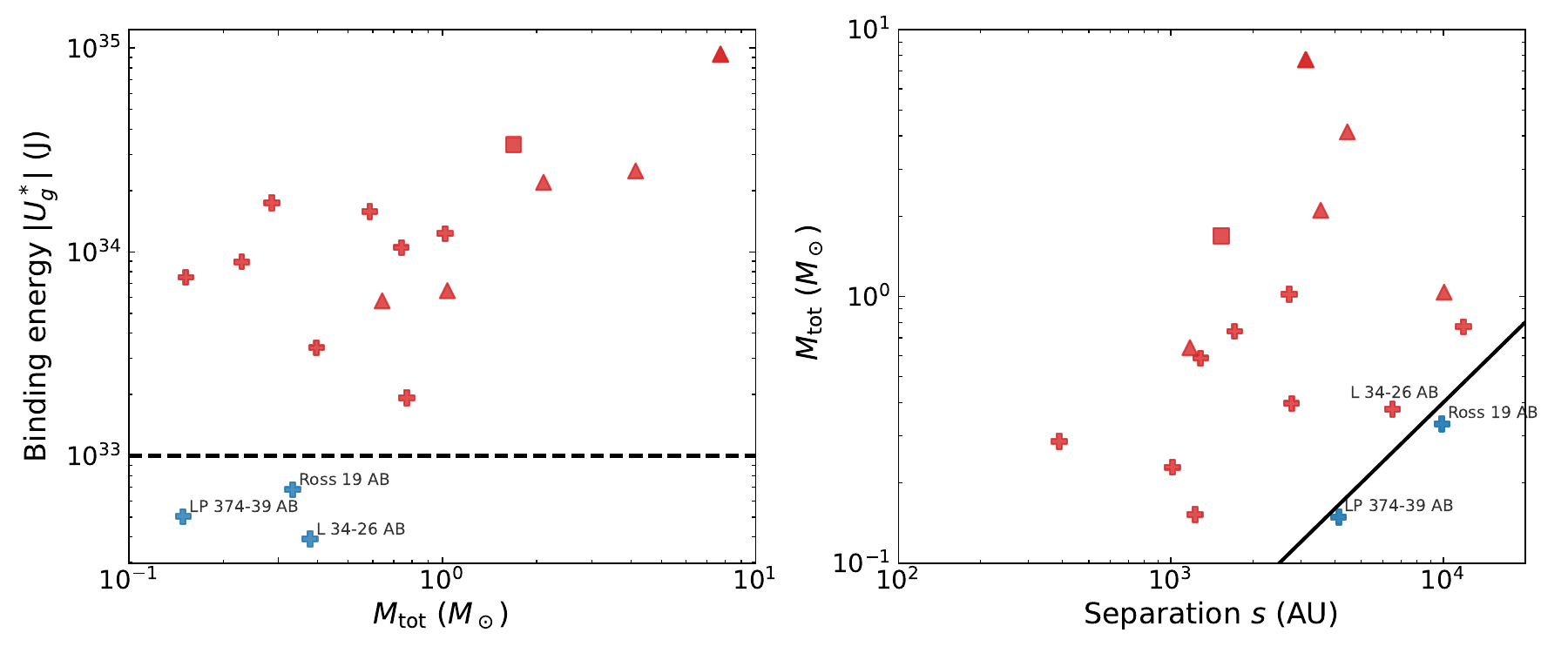}
\caption{Reduced binding energy (left panel) and projected physical separation (right panel) as a function of the total system mass. On both panels, binary systems are shown as plus symbols, triple systems as triangles, and quadruple systems as squares. In the left panel, systems highlighted in blue correspond to weakly bound systems, as defined by the binding energy threshold estimated by \citet{Caballero2010} (dashed line). On the right panel, systems also shown in blue represent fragile systems located near the 10\,Gyr diffusion limit (continuous line), taking into account stellar encounters and the estimated average binary lifetime from Eq.~(28) of \citep{Weinberg}, and also computed using the approximation from \citep{Dhital2010} (see Eq.~\ref{eq:lifetime}).}
\label{fig:Energe}
\end{figure*}

Finally, two other low-mass systems also have half-lives shorter than 1\,Gyr. The first, LP\'374-39 with companion 2MASSJ11225550+2550250 (LP374-39B), has an unknown age; this limits our interpretation, but its metallicity and Galactic location may provide additional clues regarding its dynamical history and potential vulnerability to external perturbations. The second, Ross\,19 with companion Ross\,19\,B, has an estimated age of $7.2^{+3.8}_{-3.6}$Gyr \citep{2021Schneider}, significantly older than its $\sim$1\,Gyr half-life, which, similarly to the HD~170573 system, suggests a relatively recent excitation into its current orbital configuration. Notably, Ross\,19 is the most metal-poor system in the sample of BD-hosting stars.

{  
\subsection{Long-term dynamical stability of ultra-wide companion systems}
\label{long-term}

The ultra-wide companions investigated are expected to be particularly sensitive to external perturbations and therefore provide valuable constraints on the long-term dynamical evolution of low-binding-energy binaries. The analysis presented in Sect.~\ref{sec:half-life} offers insight into the global dynamical properties of these systems and enables an assessment of their long-term stability.

To investigate their possible future evolution, we computed their gravitational binding energies, estimated the diffusion-driven disruption limit following the formalism of \citet{Weinberg} for a characteristic age of 10~Gyr, and evaluated the potential impact of perturbations induced by giant molecular clouds, galactic tides, and passing stars.

The gravitational binding energy was calculated using the expression introduced by \citet{Caballero2009}:
\begin{equation}
|U_{\mathrm{g}}^{\ast}| = \frac{G M_1 M_2}{s}\label{eq:energie},
\end{equation}
where $G$ is the gravitational constant, $M_1$ and $M_2$ are the masses of the two components, and $s$ is the projected physical separation.

In several previous studies \citep{Chinchilla2020,Gonzalez2023}, a correction factor of 1.26 has been applied to convert the projected separation $s$ into an estimate of the orbital semi-major axis $a$, under the assumption of circular orbits and isotropic inclination distributions (i.e.\ $a = 1.26\,s$). However, numerical simulations of wide binaries by \citet{Tokovinin2020} have shown that, for population-level studies, the median projected separation provides a reliable estimate of the median semi-major axis without the need for such a correction. As these arguments do not apply to individual systems, and to remain consistent with previous statistical analyses, no projection correction was applied in this work.

Over the lifetime of wide binary systems, catastrophic encounters such as direct collisions are exceedingly rare compared to the cumulative effect of weak gravitational perturbations from passing field stars \citep{Weinberg}. We therefore accounted for long-term stellar perturbations using the simplified statistical prescription introduced by \citet{Dhital2010}, which provides an estimate of the maximum separation at which a system of a given total mass can survive for a given time:
\begin{equation}
a_{\max} \approx 1.212 \, \frac{M_{\mathrm{tot}}}{t_{\ast}} \label{eq:lifetime},
\end{equation}
where $M_{\mathrm{tot}}$ is the total system mass in units of $M_{\odot}$, $t_{\ast}$ is the system age in Gyr, and $a_{\max}$ is the projected separation expressed in parsecs.

Estimating the reduced binding energy using Eq.~\ref{eq:energie} requires knowledge of the mass of each component. In our sample, some companions (ULAS\,J150457.65+053800.8, WISE\,J183207.99-540942.0) lack mass estimates in the literature, while Ross\,19B has a mass estimated between $15$ and $44~M_{\rm Jup}$ \citep{2021Schneider}.

To overcome these gaps, we combined the age PDF of each star (described in Sect.~\ref{const_map}), interpolated the ATMO2020 models \citet{atmo_2020} at the companion's M$_{W2}$ and derived a most-likely mass and 1$\sigma$ confidence interval. The most likely mass is used in MCMC simulations presented in Sect.~\ref{sec:mcmc}. For instance, for Ross\,19B, we adopted $M_{\rm adopted} = 24.1$\,M$_{\rm Jup}$, consistent with the range provided by \citep{2021Schneider}. Adopted masses for all other companions are listed in Table~\ref{tab:stellar_parameters}(see Appendix~ \ref{appendix:table}).

Figure~\ref{fig:Energe} presents the reduced binding energy (left panel) and the projected physical separation (right panel) as a function of total system mass for the ultra-wide companion sample. In the left panel, the horizontal line marks the empirical lower envelope of the most weakly bound systems, around $10^{33}$~J, first identified by \citet{Caballero2010}. The right panel shows the diffusion-driven disruption limit at 10~Gyr derived from the formalism of \citet{Weinberg}.

In both panels, the systems Ross~19 AB and LP~374--39~AB, discussed in detail in Sect.~\ref{sec:half-life}, lie close to the inferred dynamical stability boundaries, suggesting that their present configurations may be susceptible to external perturbations. In contrast, the fragile system L~34--24~AB, despite its low binding energy \citep{Zhang2021}, appears capable of surviving in its current configuration for at least several million years.

We further investigated the potential impact of perturbations induced by giant molecular clouds on wide companion systems using the formalism developed by \citet{Weinberg}, which describes the orbital evolution of binaries subjected to external gravitational perturbations.

Following \citet{Faherty2010}, we assume that the analytic Fokker--Planck coefficients derived by \citet{Weinberg} are applicable in the regime dominated by single, impulsive encounters\footnote{Assuming 
\(
G M/(a\,\epsilon V_{\mathrm{rel}}^{2}) \ll \left(M/M_{\mathrm{p}}\right)^{2}
\).}. Within this framework, we evaluate the characteristic impact parameter for interactions with GMCs, which depends on both the total system mass and the orbital separation of the companion\footnote{We adopt $V_{\mathrm{rel}} = 20~\mathrm{km\,s^{-1}}$, $\epsilon = 0.1$, $M_{\mathrm{GMC}} = 5 \times 10^{5}\,M_{\odot}$, and $M_{\mathrm{p}} = 0.7\,M_{\odot}$, following \citet{Weinberg} and \citet{Close2007}.}

In this regime, the impact parameter associated with the Fokker--Planck approximation scales as
\(
b_{\mathrm{GMC,FP}} \propto M^{-1/4} a^{3/4},
\)
while the maximum impact parameter scales as
\(
b_{\max} \propto a^{3/2} M^{-1/2},
\)
where $M$ denotes the total mass of the system and $a$ the orbital separation.

For all companions in our sample, we find that $b_{\mathrm{GMC,FP}}$ exceeds $b_{\max}$, indicating that perturbations due to giant molecular clouds are unlikely to have significantly affected these systems. For the sample as a whole, the characteristic impact parameters remain above the critical disruption threshold, suggesting that GMCs have not played a dominant role in the dynamical evolution of these wide companion systems.

}

{  
\section{MCMC Statistical Analysis}
\label{sec:mcmc}
We used a statistical framework based on the Markov Chain Monte Carlo (MCMC) sampling method described in \cite{2018Fontanive,2019Fontanive} to constrain the distributions of substellar companions in our observed sample. 
The analysis was performed using the \textsc{emcee} Python package \citep{ForemanMackey2013}, which implements the affine-invariant ensemble sampler introduced by \citet{GoodmanWeare2010}.

We use the \citet{2019Fontanive} framework and adopt the parametric power-law model introduced by \citet{Nielsen2019}, which provides a description of the underlying population of substellar companions. 
The differential occurrence rate is expressed as
\begin{equation}
\mathrm{d}n^{2} = f \,
M^{\alpha} a^{\beta}
\, \mathrm{d}M \, \mathrm{d}a ,
\label{eq:distribution}
\end{equation}
where $\alpha$ and $\beta$ are the power-law indices of the companion mass and semi-major axis distributions, respectively. 
The parameter $f$ denotes the fraction of stars hosting at least one companion within the explored parameter space and is treated as a free parameter. The occurrence rate is normalised over the interval of $5-80$\,M$_{\rm jup}$ and separations of  \numAUInfthousandau--\numAUSupthousandau\,AU. We used flat priors on all parameters: $\alpha$ and $\beta$ bounded to [-5, +5], and $f$ bounded to [10$^{-5}$, 1]. We included the geometric effect of the observed projected separation, while $dn$ is defined in terms of the three-dimensional separation. Details regarding the implementation of this correction are given in Appendix~\ref{appendix:projection}.

\label{sec:results}

To ensure robust convergence, the MCMC analysis was performed using $2 \times 10^{3}$ walkers, each evolved over $10^{3}$ steps. 
After an initial burn-in phase, the ensemble of walkers expanded from their initial positions and reached a stable and representative sampling of the parameter space, converging towards regions of maximum posterior density, similarly to the behaviour reported in \citet{2019Fontanive}.

The complete results of our MCMC analysis on the entire sample are presented in Fig.~\ref{fig:Resu}. In order to investigate the host-mass dependency of substellar companions, we repeated the analysis for the M-dwarf and FGK sub-samples (see Fig.~\ref{fig:Resu}) in Table~\ref{tab:parameters-1000au}.

\begin{figure}[htb!]
\centering
\includegraphics[width=1.0\linewidth]{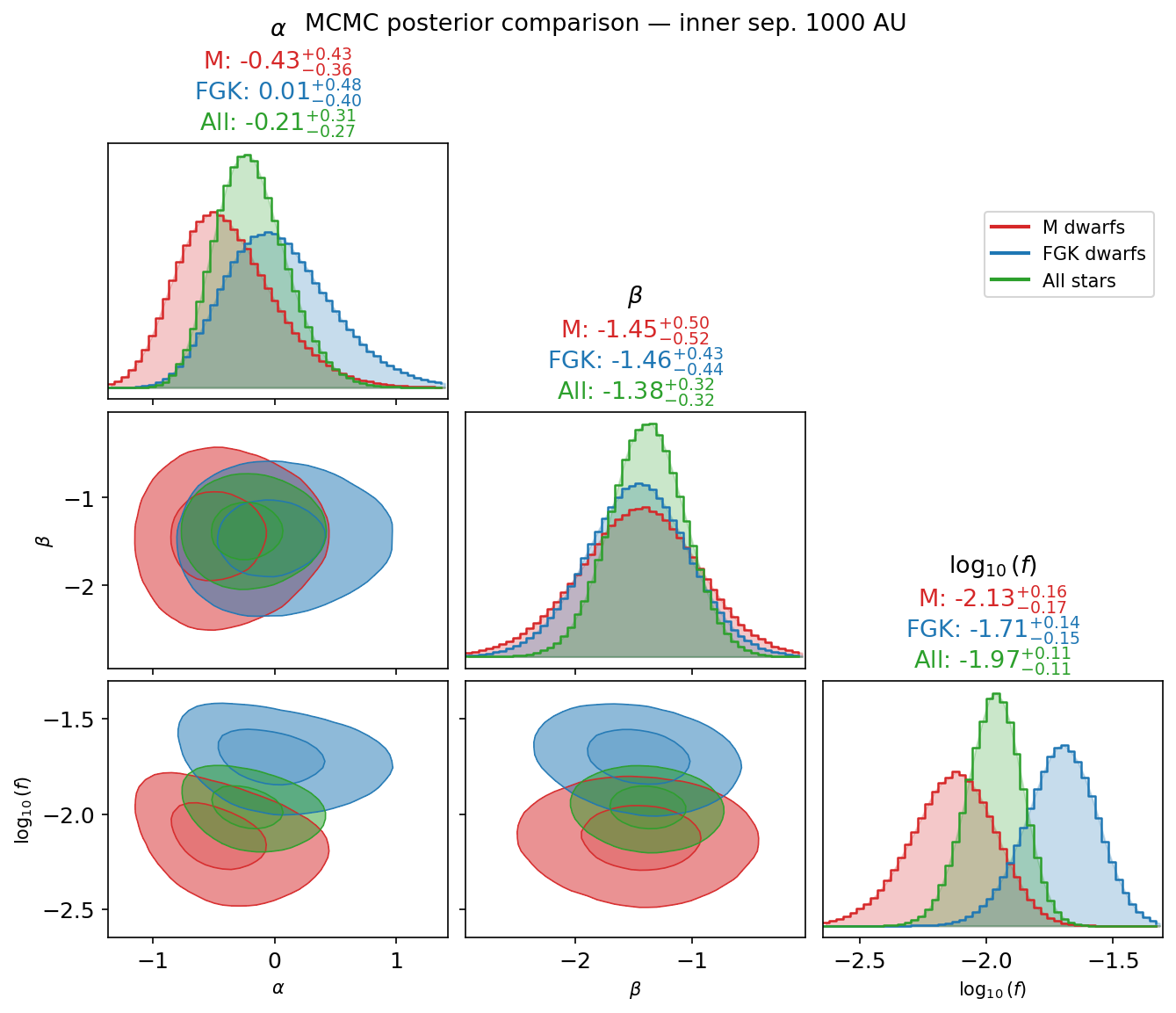}
\caption{Comparison of the MCMC posterior distributions for the parameters of the double power-law model fitted to the full sample, describing the population of giant planets with masses between 5 and 80~$M_{\mathrm{Jup}}$. The model also includes the overall occurrence rate $f$ normalised between \numAUInfthousandau\ and \numAUSupthousandau~au.
The results indicate a low logarithmic occurrence rate and negative power-law indices for the full sample, suggesting a higher frequency of planetary companions at lower masses and smaller projected separations. The inferred values are consistent with Öpik's law \citep{opik_statistical_1924}, originally proposed for stellar visual binaries and corresponding to $\beta=-1$.}
\label{fig:Resu}
\end{figure}

\begin{table}[htbp]
\caption{Best-fit parameters for companion distributions.}
\centering
\begin{tabular}{lcccc}
\hline\hline
Sample & N$_{\rm sys}$ & $\alpha$ & $\beta$ & $\log_{10} f$ \\
\hline
All & $\numallthousandau$ & $\alphaallthousandau$ & $\betaallthousandau$ & $\logfallthousandau$ \\
FGK  & $\numfgkthousandau$ & $\alphafgkthousandau$ & $\betafgkthousandau$ & $\logffgkthousandau$ \\
M  & $\nummthousandau$ & $\alphamthousandau$ & $\betamthousandau$ & $\logfmthousandau$ \\
\hline
\end{tabular}
\label{tab:parameters-1000au}
\tablefoot{
The parameters correspond to the best-fitting model derived from companions with projected separations between \numAUInfthousandau\ and \numAUSupthousandau\ AU. Quoted uncertainties correspond to the 68\,\% confidence intervals.
}
\end{table}
For the sample of all stars, the best-fit parameters obtained from the MCMC analysis are listed in Table~\ref{tab:parameters-1000au}. 
These results indicate a higher occurrence rate of companions at smaller projected separations, as suggested by the negative value of $\beta$, while the mass distribution shows a mild decrease towards higher masses, corresponding to a negative $\alpha$. 
The occurrence rate, $\log_{10}f$, is also provided in Table~\ref{tab:parameters-1000au}, allowing a direct comparison with the FGK and M-dwarf subsamples.

For the FGK subsample, the inferred separation distribution indicates a decreasing companion occurrence towards wider projected separations.

The mass distribution shows a flat distribution for FGK stars and, to 1$\sigma$, a decreasing distribution towards higher masses for M dwarf companions. We constrained the occurrence rate of LTY companions; the normalisation of the occurrence is for the \numAUInfthousandau--\numAUSupthousandau~AU domain, even if the current framework has a significant recovery rate below that inner separation (Wolf\,940\,B is recovered $<\numAUInfthousandau$\,AU). As show in Fig.\,\ref{fig:carte}, for some systems close-in M-dwarf systems, we have sensitivity at the few hundred AUs. The sensitivity at smaller separations is accounted for in the Bayesian framework of \citet{2018Fontanive} outside of the bounds of normalisation.

{
Assuming Gaussian uncertainties and independent measurements, as defined in Table~\ref{tab:parameters-1000au}, we quantified the significance of the differences between the FGK and M populations by comparing the difference between the fitted values with their combined uncertainties. The statistical significance is calculated as:

\begin{equation}
S = \frac{|X_{\rm FGK}-X_{\rm M}|}
{\sqrt{\sigma_{\rm FGK}^{2}+\sigma_{\rm M}^{2}}},
\end{equation}

where $X$ denotes the parameter being compared and $\sigma$ its associated uncertainty.

Applied to the companion occurrence rates ($\log_{10}(f)$), this comparison yields a difference at the 2.1$-\sigma$ level, corresponding to a $p$-value of $p\simeq0.04$. This result provides a marginal indication of a difference between the companion occurrence rates around FGK- and M-type stars. This suggestion of a higher occurrence rate of companions around FGK stars compared to M dwarfs is consistent with the trends reported by \citet{PROJET_WEIRD} and \citet{Nielsen2019}, although the separation ranges differ, even if the explored mass domains are comparable.

We further applied this approach to the parameters $\alpha$ and $\beta$ describing the companion mass-ratio distributions. The resulting differences are below 1$\sigma$ ($S_{\alpha}\simeq0.7$--0.8 and $S_{\beta}\simeq0.02$), indicating that these parameters are statistically consistent between the two populations. Therefore, although FGK stars exhibit a higher companion occurrence rate within the mass and separation ranges explored in this study, no statistically significant difference is found in the shape of the companion mass-ratio distributions between FGK- and M-type host stars.
}

\section{Comparison with population models}
\label{sec:comp}

We compare our measurements with population synthesis predictions for various formation scenarios, with a particular focus on gravitational instability (described in Sect.~\ref{model:GI}), which effectively spans the parameter space probed by our sample.

Our sample is dominated by massive companions at wide orbital separations. Current core accretion (CA) population synthesis models do not reach these regimes, with predicted companions extending up to $\sim$465\,AU in separation and $\sim$0.045\,\Mjup\ in mass \citep{2018Mordasini,2021Emsenhuber}. Exploring our parameter space would therefore require uncertain extrapolations.

Additional dynamical mechanisms, such as Kozai--Lidov cycles in hierarchical triple systems \citep{Naoz2016} or turbulent fragmentation of the parent molecular cloud \citep{2007Whitworth}, may also contribute to shaping the observed architectures. However, given the large companion masses and separations considered here, our analysis primarily relies on the predictions of GI models.}

The overall slope in separation distribution, $\beta=\betaallthousandau$, is nearly consistent with the \"Opik law \citep{opik_statistical_1924} for stellar binaries, where abundance is uniform in log separation ($\beta=-1$). There is no significant difference between M dwarfs and FGK stars in the separation power law (respectively $\betamthousandau$ and $\betafgkthousandau$). 

\subsection{Gravitational Instability}
\label{model:GI}

The GI population synthesis framework was developed by \citet{2013Forgan} and updated by \citet{2018Forgan}. This rapid formation mechanism operates in massive, cold protoplanetary discs ($Q<1.5$), which become gravitationally unstable and fragment into self-gravitating clumps. The initial fragments, with masses ranging from 3 to 125~M$_{\rm Jup}$, subsequently evolve through a \textit{tidal downsizing} process involving contraction, orbital migration, gas and dust accretion, and tidal disruption. This evolution produces gas giant planets, brown dwarfs, or low-mass stellar companions, predominantly at wide orbital separations ($>$20~AU).

Host star masses $M_*$ range from 0.8 to 1.2~M$_\odot$ (extendable to 0.5--2~M$_\odot$), with high disc-to-star mass ratios ($q=0.25$--1.0) favouring fragmentation beyond 30--40~AU. The improvements introduced by \citet{2018Forgan}, including fragment--fragment interactions and refined migration prescriptions (Types I/II), reduce the excess of close-in objects and improve agreement with observations.

The Disc Instability Population SYnthesis \citep[DIPSY; ][]{2025SchibI} framework presents a new global population synthesis model for companion formation via gravitational disc instability. This framework solves the 1D viscous disc evolution with infall from the molecular cloud core, stellar irradiation, photoevaporation, and self-gravity fragmentation, covering companions from planetary to low-stellar masses around host stars ranging from brown dwarfs to massive B-type stars, including M dwarfs~\citep{2025SchibII}.

In baseline populations, $\sim10$\,\% of discs fragment, yielding mostly brown dwarfs (75\,\%) at large separations ($>100$~au), with strong sensitivity to interconnected processes such as gas accretion, orbital migration, and N-body interactions~\citep{2025SchibII}.

The DIPSY model series explore the impact of fundamental physical assumptions on companion formation through gravitational instability. DIPSY-0 establishes the fiducial reference model, characterised by disc-regulated gas accretion, a restricted number of initial fragments, and protoplanetary disc parameters consistent with observational constraints. DIPSY-1 tests the effect of unconstrained gas accretion, allowing higher mass consumption rates. DIPSY-2 removes accretion entirely while increasing initial fragment masses to preserve global statistical properties. DIPSY-3 simplifies initial orbital conditions by imposing zero eccentricities, eliminating complex early-stage dynamics. DIPSY-4 increases the radial extension of the protoplanetary disc to assess the effect of available fragmentation area. DIPSY-5 reduces the number of fragments produced per instability event, testing sensitivity to late-stage gravitational interactions and collisions. DIPSY-6 explores variations in disc viscosity and photoevaporation parameters, two critical processes governing dissipation and thermal evolution. Despite this extensive range of parametric variations spanning distinct aspects of disc physics and initial conditions, the final statistical distributions of companion mass and separation remain remarkably similar across all models. This result demonstrates the robustness of the gravitational instability scenario and indicates that gravitational dynamical processes (mutual gravitational interactions, inelastic collisions, and orbital migration) dominate system evolution independently of these initial physical details.

\begin{figure}[htb!]
\centering
\includegraphics[width=1.\linewidth]{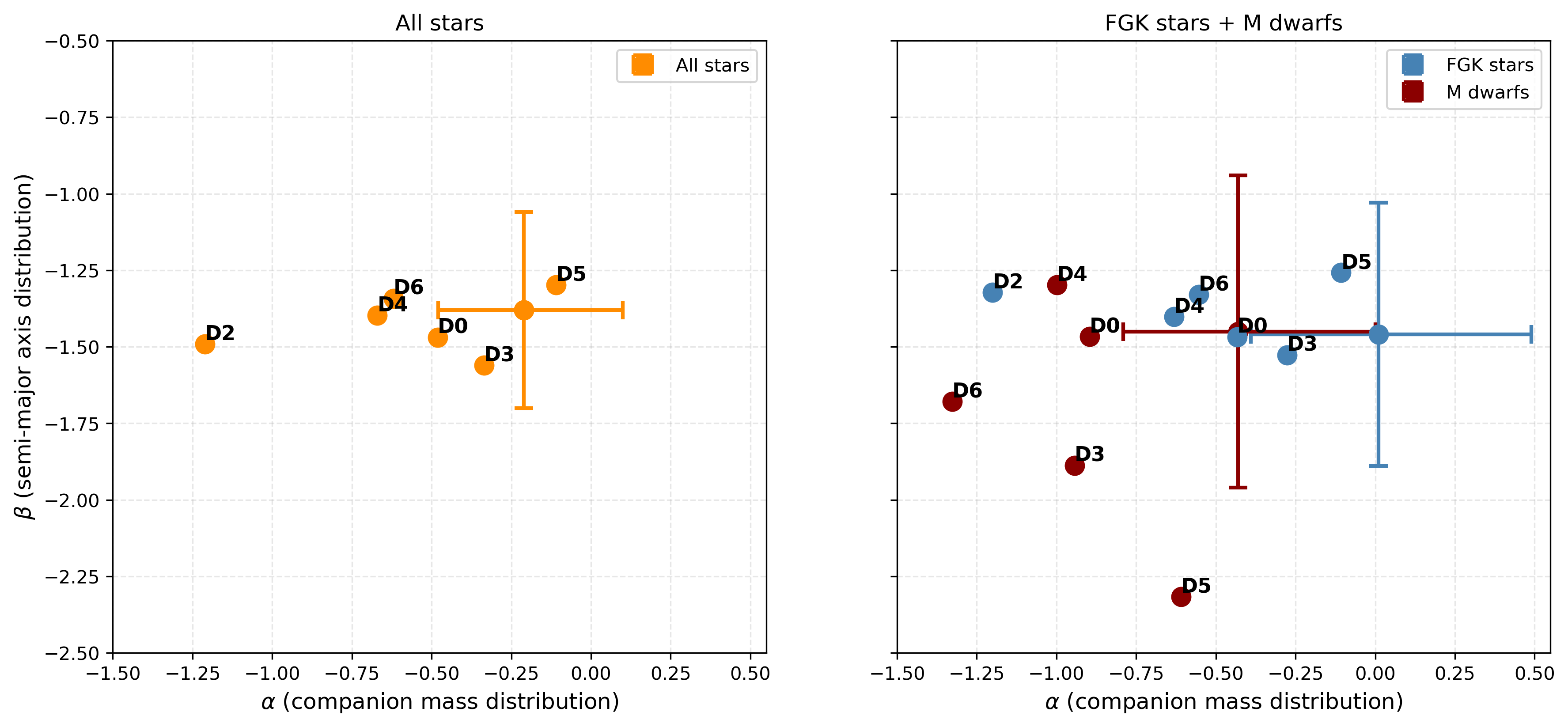}
\caption{Empirical $\alpha$ and $\beta$ slopes for the M-dwarf and FGK samples ( right panels), and for the full sample (left panels) compared to the DIPSY \citep{2025SchibII} models (D0 to D6, corresponding to versions DIPSY-0 to DIPSY-6). The FGK companion distribution is consistent at the 2$\sigma$ level with all but the DIPSY-2 population. M dwarf DIPSY companions tend to have much steeper mass and semi-major axis distributions than observed.}
\label{fig:gi_model}
\end{figure}
}

We compared the constraints on the parameters $\alpha$ and $\beta$ derived from our different samples (the full sample, FGK dwarfs, and M dwarfs) with the predictions of the DIPSY model\footnote{\url{https://zenodo.org/records/17380339}} for companions with masses between $0.5$ and $80\,M_{\mathrm{Jup}}$ and projected separations ranging from \numAUInfthousandau\ to \numAUSupthousandau\,AU. Figure~\ref{fig:gi_model} shows this comparison, with the left panels corresponding to the full sample and the right panels to the FGK- and M-dwarf subsamples.

Overall, the DIPSY models provide a good match to the constraints derived from the full sample. Except for the DIPSY-2 model, all predictions remain consistent with the observational constraints at the 2$\sigma$ level. This agreement suggests that the formation and dynamical evolution scenarios implemented in these simulations are broadly capable of reproducing the observed distribution of wide-orbit companions. In particular, the results favour a population of relatively massive companions, including high-mass brown dwarfs, at small orbital separations.

We stress that the DIPSY models are initial mass/separation distribution, while the observed sample has, in most cases, had many Gyr of interaction with field stars and giant molecular clouds. It is not a primeval distribution and the correspondence with models should be seen as indicative until these effects are included in a physically consistent framework.

\subsection{Comparison with IMF measurements in the sub-stellar regime}

A comparison with the initial mass function (IMF) is only meaningful under the assumption that companion masses are themselves drawn from the IMF, which is not trivial. This assumption is approximately valid for stellar binaries at large separations around massive primaries \citep{Moe2017}. 
{Nevertheless, such a comparison should be interpreted with caution, since low-mass companions located at very large separations ($>1000$ au) may have formed through turbulent fragmentation of the parental molecular cloud rather than through processes associated with the circumstellar disc of the primary star.}

{Most IMF studies define the mass function as \hbox{$dN/dM \propto M^{-\alpha}$}, whereas we adopt the convention used in exoplanet and sub-stellar companion studies \citep[$dN/dM \propto M^{\alpha}$; ][]{Cumming2008,Nielsen2019,baron_2018}. The sign of the slope must be reversed when comparing our results with published IMF measurements.}
For young associations, the reported IMF slopes are $\alpha = 0.33 \pm 0.19$ over $0.01$--$0.1\,M_\odot$ \citealt{Muzik2025} and $\alpha = 0.14 \pm 0.19$ over $0.01$--$0.2\,M_\odot$ \citealt{Rom2026}. The field IMF within 20\,pc also yields $\alpha = 0.6$ over the mass range $0.01$--$0.05\,M_\odot$ \citep{kirkpatrick_2024}. 
{Taking into account this difference in slope convention, our companion mass functions are consistent with these IMF at the $<2-\sigma$ level.}

\section{Prospects\label{perspectives}}

We performed a statistical analysis to constrain the distribution of substellar companions in the planetary mass regime. This study is based on a compilation of direct imaging surveys from the CatWISE archives and assumes a power-law distribution for substellar companion masses and {  projected separation}.

\subsection{Expanding on direct imaging surveys}

{  
The mass distribution derived in this study is broadly consistent with that reported by \citet{PROJET_WEIRD}, despite notable differences in methodology and sample selection criteria. This agreement indicates that the overall trends identified in both studies are relatively robust against observational biases and methodological differences. In particular, \citet{PROJET_WEIRD} performed a statistical analysis combining high‑resolution direct imaging and deep seeing‑limited observations to simultaneously probe the full orbital separation domain from 5 to 5000~AU around a sample of 344 young, nearby stars. By assuming that the companion distribution in mass and semi‑major axis follows a power‑law, they constrained the parameters of this distribution and obtained an overall occurrence rate of $\sim$11~\% for companions with masses between 1 and 20~$M_{\mathrm{Jup}}$ over these separations, finding that companion frequency decreases with increasing semi‑major axis and, marginally, with companion mass, while increasing with host star mass. In this context, our work revisits Gyr‑old analogues of these systems; therefore, the agreement observed between the inferred mass distributions is not unexpected.

In contrast, our results show significant discrepancies with those presented by \citet{Nielsen2019}. 
These differences may arise from variations in the assumptions adopted for population modeling, or from intrinsic differences in the parameter space explored rather than from the properties of the samples themselves.

Specifically, \citet{Nielsen2019} analysed the demographics of giant planets and brown dwarfs around approximately 300 stars, at orbital separations between 10 and 100~AU. They found an occurrence rate of roughly 9~\% for planets of 5–13\,$M_{\mathrm{Jup}}$ and about 0.8~\% for brown dwarfs of 13–80\,$M_{\mathrm{Jup}}$ within this separation range.
Our WISE-based observations probe significantly wider orbital separations and different mass regimes. It is therefore not surprising that the observed populations differ, as they sample distinct regions of the parameter space that are dominated by different formation mechanisms.

\subsection{Origin of substellar companions}

The trend highlighted by our study remains qualitatively compatible with expectations from disc GI, in which more massive discs around higher-mass stars are generally expected to be more prone to fragmentation. However, the comparison with the DIPSY predictions, particularly in the M-dwarf regime, indicates that the agreement is not uniformly supported across the full stellar-mass range and should therefore be interpreted with caution.

Alternative formation pathways may also contribute to the population of wide-orbit substellar companions. In particular, turbulent fragmentation within the parental molecular cloud could in principle produce objects in the planetary-mass regime, in a manner analogous to the formation of very low-mass stars or brown dwarfs. Recent observational studies probing the lowest-mass end of the initial mass function and the population of very low-mass brown dwarfs and planetary-mass objects have begun to explore this regime \citep{DeFurio2025,Luhman2025}.

However, current theoretical and observational studies do not yet provide quantitative predictions for the occurrence rate, orbital properties, or mass distributions of bound companions comparable to those available for disc GI. As a result, it remains difficult at present to directly confront this scenario with the observational demographics of substellar companions.

Future theoretical work, including numerical simulations capable of producing testable predictions, will be required to assess the relative contribution of this mechanism. Expanding the sample of well-characterised systems will also help clarify the origin of these objects.
}

\subsection{Understanding weakly-bound binaries}

Interestingly, some of the detected or candidate systems appear dynamically fragile given their estimated ages. 
 Although this apparent discrepancy may be compatible with statistical survival in rare instances (some systems are at $>10$ half-lives), it likely indicate that the companions have not remained on their current wide orbits throughout the entire evolutionary history of the system. Instead, dynamical processes migrated these companions outward to their present fragile configurations. Understanding these processes is beyond the scope of the current work, but the fact that two out of 17 systems are in that state suggests that this is a common behaviour. Future $N$-body simulations incorporating realistic Galactic models, such as the Besançon model \citep{Robin2003,Czekaj2014}, together with the Galactic potential and stellar encounters, will be required to quantify the long-term dynamical evolution of these systems and to determine whether their present-day configurations can naturally arise from late-time dynamical processes.
 
\subsection{Reaching lower masses at wide separations}
The decision to limit our sample to a 20\,pc radius was driven by the goal of maximising both statistical completeness and depth, rather than extending the survey to larger but less reliable distances. Beyond this limit, especially for the coolest substellar objects such as T and Y dwarfs, we approach the sensitivity threshold of the WISE survey in the $W2$ band. This constraint significantly hampers the reliability of detections and the completeness of the sample at faint absolute magnitudes (M$_{W2}$), thereby reducing the number of detectable objects.

In principle, these limitations could be mitigated through improved angular resolution, which would enable the combination of direct imaging with astrometric and radial velocity measurements. However, no upcoming mission operating near 4~$\mu$m is expected to offer better angular resolution than WISE. Although the James Webb Space Telescope (JWST; \citealt{2023Rigby}) provides vastly superior sensitivity—by several orders of magnitude—its extremely limited sky coverage restricts its usefulness for large-scale statistical studies. Similarly, the SPHEREx mission \citet{2020Crill}, which will offer full-sky spectral coverage between 0.75 and 5\,$\mu$m, is limited by its relatively coarse angular resolution of about 6$^{\prime\prime}$, similar to that of WISE. \\
Wide-field missions such as \textit{Euclid}
(\citealt{2011Laureijs}; \citealt{2022EuclidCollab}) and the
\textit{Nancy Grace Roman Space Telescope}
(\citealt{2015Spergel}; \citealt{2019Akeson}) will 
provide an important complement to these surveys, mostly for L and T dwarfs and to a lesser extent for Y dwarfs. With sky coverage
approaching one third of the sky and angular resolutions of
$\sim0.1{-}0.2''$, they will enable efficient detections of brown
dwarfs and low-mass companions across large stellar samples, although
their sensitivity remains less favourable for probing the coldest
objects than surveys operating near 4~$\mu$m. The recently commissioned Vera C. Rubin Observatory will further strengthen this observational landscape. Through the Legacy Survey of Space and Time (LSST), it will provide deep multi-epoch imaging in the red optical bands over approximately 18\,000~deg$^2$, corresponding to  nearly half the sky \citep{Ivezic2019}. Its unprecedented combination of depth, sky coverage, and astrometric precision will enable the identification of large samples of substellar objects and wide substellar companions through common proper motion searches, thereby providing robust constraints on the demographics of the substellar population, particularly for the warmer brown dwarfs that remain detectable at optical wavelengths \citep{LSSTScienceBook2009,2022Gizis}. The statistical power of these samples will complement mid-infrared surveys by probing a larger sample of host stars for L and T companions.

\begin{acknowledgements}
The authors thank the anonymous referee for the constructive comments and suggestions that greatly improved the overall quality of the paper.

This research was supported by the Laboratory of Physics and Environmental Chemistry (LPCE) at Universit\'e Joseph KI-ZERBO in Burkina Faso. It also received financial support through a research mobility grant awarded for a research stay at the Universit\'e de Montr\'eal, within the Trottier Institute for Research on Exoplanets (iREx), thanks to the Canadian government via Global Affairs Canada.\\
SO is grateful to the \textsc{Erasmus} programme for financial support,
provided through the Universit\'e C\^ote d’Azur, the Universit\'e Joseph KI-ZERBO,
and the Observatoire de la C\^ote d’Azur / Laboratoire Lagrange in Nice, France\\
\'EA, NJC \& RD acknowledge the financial support of the FRQ-NT through the Centre de recherche en astrophysique du Qu\'ebec as well as the support from the Trottier Family Foundation and the Trottier Institute for Research on Exoplanets.\\
\'EA \& RD acknowledge support from the Canada Foundation for Innovation (CFI) programme, the Universit\'e de Montr\'eal and Universit\'e Laval, the Canada Economic Development (CED) programme and the Ministere of Economy, Innovation and Energy (MEIE).

This publication makes use of data from the WISE (Wide-field Infrared Survey Explorer) mission, a joint project funded by NASA, which enabled sky mapping at 20 different epochs over approximately 12 years. It also uses the VizieR catalogue service and the SIMBAD database, both operated by the Centre de Données Astronomiques de Strasbourg (CDS), France.

This work also uses data from the Gaia mission of the European Space Agency (ESA), accessible at \url{https://www.cosmos.esa.int/gaia}, and processed by the Gaia Data Processing and Analysis Consortium (DPAC). DPAC is funded by national institutions, particularly those participating in the Gaia Multilateral Agreement.
\end{acknowledgements}

\bibliography{References.bib}
\bibliographystyle{aasjournal.bst}

\begin{appendix}
\section{Properties of recovered companions}
\label{appendix:recovered}
We present here the properties of the companions recovered with the current framework. The main parameters are compiled in Table~\ref{tab:stellar_parameters}.
\subsubsection*{BD+06 2986~AB}
The BD+06 2986 system was identified as a common proper motion pair by \citet{2010Scholz}. The primary component, a solar-type dwarf detected in Gaia DR3, has a metallicity of \hbox{[Fe/H] = $-0.37$} \citep{2010Scholz,2025Burgasser}. The companion, BD+06 2986B (also known as HIP 73786B or ULAS J150457.65+053800.8), shares a consistent proper motion with the primary, confirming their physical association. Its effective temperature is estimated at 953 $\pm 48$\,K \citep{2025Burgasser}.

\subsubsection*{BD+13 2618~ABC}
BD+13 2618 (Ross 458) is a hierarchical triple system consisting of a close binary, Ross 458~AB, and a distant substellar companion, Ross 458 C \citep{2010Burgasser}. The primary system, detectable in Gaia DR3, exhibits a supersolar metallicity with ${\rm [Fe/H]} = +0.20\pm 0.05$ and an estimated age between 150 and 800\,Myr \citep{2010Burgasser}. BD+13 2618~C (Ross 458~C) shares a common proper motion with the central binary at a wide projected separation and has an effective temperature of 790 $_{-10}^{+63}$\,K \citep{2025Burgasser}. 

 \subsubsection*{BD+21 2486~ABC}

BD+21\,2486 is a triple system consisting of a close stellar binary, BD+21\,2486\,AB, unresolved in Gaia DR3 , and a wide-separation substellar companion, 2MASS\,J13054106+2046394 (also known as GJ\,499C), which is detected in Gaia DR3. The companion shares a common proper motion with the binary, indicating a physical association \citep{2003Cruz, 2013Gomes}. The system is estimated to be between 3 and 5\,Gyr old \citep{2024Calamari}. The effective temperature of the substellar companion has been reported as \hbox{$1574\pm170$\,K} \citep{2013Gomes} and 1603\,K \citep{Kirkpatrick_2019}. 

\subsubsection*{G 204-39~AB}
G 204-39 is a binary system identified by \citet{2010Faherty}, composed of a host star detected in Gaia DR3 and a substellar companion. The estimated age of the primary star ranges between 0.5 and 1.5\,Gyr. Both objects share a common proper motion, indicating likely gravitational binding despite a large angular separation. The companion, G 204-39B, is a BD with an effective temperature estimated at $901 \pm 88$\,K \citep{Kirkpatrick_2019}.

\subsubsection*{G 48-43\,AB}

LHS\,6176 is a binary system composed of a low-mass dwarf star, G 48-43 (LHS\,6176), detected in Gaia DR3, and a substellar companion, G 48-43 B (WISE J095047.32+011733.3), discovered by \citet{2013Burningham}. The primary component is a low-mass dwarf with a metallicity of [Fe/H] = $-0.30 \pm 0.1$ and an age greater than 3.5 Gyr \citep{2013Burningham}. The companion, WISE J095047.32+011733.3, shares common proper motion with the host star and is widely separated. It has a metallicity of ${\rm [M/H]} = -0.21$\,dex and an effective temperature of \hbox{$802 \pm 64$\,K }\citep{2025Burgasser}.

\subsubsection*{Gl 570\,ABCD}
The Gl 570 system is a hierarchical configuration composed of four components. Gl\,570\,A is the primary star, while Gl\,570\,B and Gl\,570\,C form a close binary subsystem. The fourth component, Gl\,570\,D, is a BD with an estimated effective temperature of 901\,$\pm$\,88\,K \citep{Kirkpatrick_2019}. 

\subsubsection*{HD 170573 \& WISE J183207.99-540942.0}
HD 170573 is a solar-type star listed in Gaia DR3, with an estimated age between 9 and 13.5 Gyr. It belongs to a binary system that includes a substellar companion, WISE J183207.99-540942.0, discovered through the Backyard Worlds: Planet 9 programme \citep{2024_Rothermich}. The companion was identified based on its common proper motion with the primary star, despite a wide angular separation. Its effective temperature is estimated at 819 $\pm$ 79 K.

\subsubsection*{HD 65486\,AB}
HD 65486 (HIP 38939) is a solar-type dwarf star detectable in Gaia DR3, slightly metal-poor with \hbox{[Fe/H]$ = -0.24$}, and an estimated age of $900^{+1900}_{-600}$ Myr \citep{2012Deacon}. The star forms a wide binary system with a substellar companion, HIP 38939 b, with an effective temperature of $1095\pm88$\,K \citep{Kirkpatrick_2019}, detected through spectroscopic observations and discovered by \citet{2012Deacon}.

\subsection*{HD 126053 AB}
HD 126053 (BD +01 2920, HIP 70319) is a binary system consisting of a solar-type primary star and a substellar companion, discovered by \citet{2012Pinfield}. The primary star is a slightly metal-poor dwarf, identified in Gaia DR3, with an iron abundance of [Fe/H] = $-0.38 \pm 0.06$ and an estimated age ranging from 2.3 to 14.4\,Gyr \citep{2024Calamari}. The system hosts a distant companion, WISE J142320.84+011638.0, which shares a common proper motion with the host star, suggesting a physical association. This companion has been identified as a BD based on its photometric and spectroscopic properties, with an effective temperature of 670 $\pm$ 34 K and a metallicity of ${\rm [M/H]} = –0.35$\,dex \citep{2025Burgasser}.

\subsubsection*{ L~34-26 ~AB}
The L 34-26 (COCONUTS-2) system, composed of the primary star COCONUTS-2A and its substellar companion COCONUTS-2B, was discovered and characterised by \citet{2021Zhangz} as part of the COol Companions ON Ultrawide orbiTS (COCONUTS) programme—a wide-field imaging survey designed to detect cold and widely separated companions around nearby stars.
The two components exhibit common proper motion. COCONUTS-2A (L\,34-26) is a young, low-mass star listed in Gaia DR3, with an estimated metallicity of [Fe/H] = $0.00\pm0.08$\,dex. Its age is estimated to lie between 150 and 800\,Myr based on multiple youth indicators \citep{2021Zhangz}. The companion, COCONUTS-2B (L\, 34-26 B), is an extremely cold substellar object with an estimated effective temperature of $483 ^{+44}_{-53}$\,K \citet{2025Zhang}, placing it among the coldest planetary-mass companions known.

\subsubsection*{LP 738 -14~AB}

LP 738-14 is a low-mass dwarf star with a metallicity comparable to that of the Sun and exhibits no detectable H$\alpha$ emission. Based on its metallicity and kinematic properties, the age of the system is constrained to an upper limit of approximately 10 Gyr \citep{2012Deacon}.
It is part of a binary system that includes a substellar companion, LP 738-14B (also known as LHS 2803 b), discovered by \citet{2012Deacon}. The companion was identified based on its common proper motion with the primary star at a wide angular separation. LP 738-14B has an estimated effective temperature of 939\,$\pm$88\,K \citep{Kirkpatrick_2019}. 

\subsubsection*{LP 374-39~AB}
LP 374-39 (GJ 3657, LHS 302) is a late-type dwarf star with a metallicity of -0.1 dex \citep{2019Kuznetsov}, forming a widely separated binary system with WISEPC J112254.73+255021.5, a confirmed substellar companion discovered by \citet{Kirkpatrick_2011}. Both objects share a common proper motion, indicating a likely physical association. The companion has an estimated effective temperature of $855 \pm 88$\,K \citep{Kirkpatrick_2019}, while the age of the system remains unknown at this time.

\subsubsection*{Ross 19~AB}
Ross 19 is a low-mass dwarf star detected in Gaia DR3, with an estimated age of $7.2^{+3.8}_{-3.8}$ Gyr and a metallicity of [Fe/H] = $-0.40 \pm 0.12$ \citep{2021Schneider}. It forms a wide-separation binary system, as reported in the same study, with a very cold substellar companion, Ross 19 B, whose effective temperature is estimated to be $500_{-100}^{+115}$ K.
\subsubsection*{$\xi$ UMa AB \& WISE J111838.70+312537.9}
WISE J111838.70+312537.9 shares a common proper motion with the nearby quadruple star system $\xi$ Ursae Majoris, discovered by \citet{2013Wright}. This companion orbits the system at a wide separation. The primary star, unresolved in the Gaia DR3 catalogue \citep{gaia_dr3}, exhibits solar metallicity, suggesting a minimum system age of 2 Gyr. The effective temperature of \hbox{WISE J111838.70+312537.9} is estimated at \hbox{559 $\pm$ 88\,K} \citep{kirkpatrick_2024}.

\subsubsection*{Wolf 940~AB}
Wolf 940 is a binary system composed of the star Gaia DR3 Wolf 940 A (also known as LHS 3708 and GJ 1263), with an estimated metallicity of [Fe/H]= -0.06 $\pm$ 0.20, and a substellar companion, Wolf 940 B, discovered by \citet{2009Burningham}. The system exhibits common proper motion and a wide angular separation between the two components. The companion, Wolf 940 B, has an effective temperature of 586 $\pm$ 35 K \citep{2025Burgasser}.

\subsubsection*{Wolf 1130~ABC}
Wolf 1130 is a hierarchical triple system composed of a close binary (Wolf 1130 AB) and a distant substellar companion, Wolf 1130 C (WISE J200520.38+542433.9), discovered by \citet{2018Mace}. The primary component, Wolf 1130 A, is a low-mass star detected in Gaia DR3, with a subsolar metallicity of [Fe/H] = -0.7 $\pm$ 0.12 dex \citep{2018Mace}. It forms a tight pair with a compact white dwarf. Wolf 1130 B has an initial mass estimated at about 7.2\,\Msun \citep{kirkpatrick_2024}. The high mass of the white dwarf and the short orbital separation suggest a history of close binary evolution \citep{2018Mace}.
The third component, Wolf 1130 C, shares common proper motion with the inner binary and lies at a wide projected separation. This substellar companion has an estimated effective temperature of 644 $\pm$ 32 K and a metallicity of ${\rm [M/H]} = -0.65^{+0.10}_{-0.07}$ dex \citep{2025Burgasser}.

\subsubsection*{$\eta$ CrB\,ABC}
$\eta$ Coronae Borealis is a triple stellar system consisting of a solar-type binary, $\eta$ CrB\,AB, with an estimated age between 1.0 - 2.5\,Gyr, and a substellar companion, 2MASSW\,J1523226+301456, discovered by \citet{2001Kirkpatrick}. The companion shares common proper motion with the system and lies at a wide separation from the primary binary. Its effective temperature is estimated to be $295 \pm 76$\,K \citep{Kirkpatrick_2019}.

\subsection{Properties of known companions not recovered}\label{appendix:not recovered}
This section gathers the properties of companions previously reported in the literature around stars in our sample, but not recovered by our approach. We also discuss the possible reasons for their non-detection.

\subsection*{15 Sge~AB}
15~Sge~AB is a binary system first identified by \citet{2012Crepp}. The primary component, 15~Sge~A, is a G0V dwarf star imaged directly by \citet{2002Liu}, located at a distance of $17.77 \pm 0.014$~pc \citep{gaia_dr3}. The star shows a mild metal enrichment with a metallicity of $[{\rm Fe/H}] = 0.05 \pm 0.07$~dex, an estimated age of $2.5 \pm 1.8$\,Gyr, an effective temperature of $T_{\rm eff} = 5883 \pm 59$\,K, and a mass of $1.08 \pm 0.04\,\Msun$ \citep{2012Crepp}. An alternative stellar mass estimate of $0.960\,\Msun$ has also been reported by \citet{kirkpatrick_2024}.

The secondary component, 15~Sge~B, is an L4 dwarf orbiting at an angular separation of $1.03''$ ($18.3^{+0.4}_{-0.5}$\,AU) \citep{2012Crepp}, unresolved in WISE imaging. The companion has an effective temperature of \hbox{${\rm T}_{\rm eff} = 1533 \pm 88$\,K} \citep{Kirkpatrick_2019} and a dynamical mass of $68.7^{+2.4}_{-3.1}$~\Mjup. Its orbit is eccentric ($e = 0.50^{+0.01}_{-0.01}$), with an orbital period of about 73.3\,yr around the host star \citep{2012Crepp}.

\subsubsection*{54 Psc AB}
54 Psc (AB) is a binary system with an estimated age of $7 \pm 3$\,Gyr \citep{epsilon_2003}, composed of a K0.5V primary, 54 Psc A (HD 3651), located at a distance of $11.11 \pm 0.048$\,pc \citep{gaia_dr3}, a mass of $0.900 \pm 0.111$\,\Msun \citep{kirkpatrick_2024}, an effective temperature of $5\,203\pm23$\,K, and a metallicity of $0.14 \pm 0.02$\,dex \citep{Soubiran_2022}.

The second component, 54\,Psc\,B (HD\,3651\,B), is a \hbox{T$7.5\pm0.5$} spectral-type BD sharing a common proper motion with its host star and located at a projected separation of \hbox{$476 \pm 6$\,AU}. This cold companion has an estimated mass of \hbox{$0.051 \pm 0.014$\,\Msun} and an effective temperature of $810 \pm 50$\,K \citep{epsilon_2003}. It is undetectable in WISE observations due to its angular proximity to the primary component.

\subsubsection*{HN Peg~AB}
HN Peg A is a G0V dwarf located at 18.13\,pc \citep{gaia_dr3}. Its mass is estimated to be $1.080\pm0.140$\,\Msun \citep{kirkpatrick_2024} and its atmospheric parameters are \hbox{T$_{\text{eff}}$ = 5939\,$\pm$\,21\,K} with a metallicity of \hbox{[Fe/H] = $-0.06\pm0.02$\,dex} \citep{Soubiran_2022}. HN Peg A forms a binary system with the BD HN Peg B, which has a spectral type of T2.5 and is located at a projected separation of approximately $795\pm15$\,AU. The mass of HN Peg B is estimated to be \hbox{$0.021\pm0.009$\,\Msun} \citep{epsilon_2003}, and its effective temperature is $1043\pm23$\,K \citep{Kirkpatrick_2019}. The system's age is about $0.3\pm0.2$\,Gyr \citep{epsilon_2003}. Considering its brightness and proper motion, HN Peg B would be readily retrieved as a co-moving companion were it not for the proximity of its host star (see Figure~\ref{fig:casehnpeg}).

\subsubsection*{GJ 229~AB }
GJ 229 is an M1.5V star located at 5.76\,pc \citep{gaia_dr3}, with a mass of $0.579\pm0.007$\,\Msun, an effective temperature of $3738\pm101$\,K, and a metallicity of [Fe/H] = $-0.07\pm0.13$\,dex \citep{Antoniadis_2024}. The star has a BD companion detected through direct imaging \citep{nakajima1995discovery}, with an estimated mass of $71.4\pm0.6$\,\Mjup, a separation of 33\,AU \citep{Brandt_2021}, and a temperature of $927\pm77$\,K \citep{Kirkpatrick_2019}. Recently, analyses conducted in \citet{Xuan_2024} concluded that GJ 229~AB is a binary system. At a projected separation of $\sim$$7^{\prime\prime}$, the system would be barely resolved by WISE and, at the contrast ratio considered, is undetected. 

\subsubsection*{HD 182488 AB}
HD 182488 is a dwarf star of spectral type K0V C, with a mass of $0.940\pm0.220$\,\Msun \citep{kirkpatrick_2024}, and $0.93 \pm 0.03$\,\Msun \citep{Feng_2022}, an effective temperature of T$_{\text{eff}}$ = $5405\pm15$\,K, a metallicity of $0.19\pm0.02$\,dex \citep{Soubiran_2022}, and located at 15.607\,pc \citep{gaia_dr3}. This star, GJ\,758 (HD\,182488), hosts a cold companion, HD\,182488 B, a T9 BD, whose mass is estimated at $38.0\pm0.8$\,\Mjup \citep{Brandt_2021} and temperature at $581\pm88$\,K \citep{Kirkpatrick_2019}. The projected separation between the primary and its companion is $29.7^{+4.2}_{-5.3}$\,AU \citep{Brandt_2021}. The system is $7.5^{+1.8}_{-1.4}$\,Gyr old \citep{Brandt_2021}. At a projected separation of $\sim$$2^{\prime\prime}$, the system is not resolved by the WISE imaging.

\subsection*{SCR J1845-6357~AB}
SCR 1845$-$6357 is a nearby binary system located at a distance of 
$4.005 \pm 0.002$~pc \citep{gaia_dr3}. 
The primary component, discovered in 2004 in the \textit{SuperCOSMOS} survey 
\citep{2004Hambly}, is an M8.5V dwarf. Its mass is estimated at 
$0.087 \pm 0.009\,\Msun$ \citep{kirkpatrick_2024}, with an effective temperature of 
 2600\,K\citep{2007Kasper}. 
 
The secondary component, SCR 1845B, is a T6 \citep{2006Biller,2007Kasper} at a projected separation from the host star of only $1.170'' \pm 0.003$, corresponding to about 4.5~AU. At such a small angular 
distance, the system is unresolved in \textit{WISE} imaging. 
Evolutionary models suggest that SCR~1845B has a mass between 40 and 
50\,${\rm M}_{\rm Jup}$, with an estimated age of 1.8--3.1\,Gyr \citep{2007Kasper}. 
Its effective temperature is \hbox{$969 \pm 88$\,K} \citep{Kirkpatrick_2019}.

\subsubsection*{WISE J072003.20-084651.2~AB}

WISE~J072003.20$-$084651.2 is a binary system composed of a primary star, WISE~J072003.20$-$084651.2~A, classified as an M9.5V dwarf, located at a distance of $6.80^{+0.05}_{-0.06}$\,pc \citep{2019_Dupuy}. The primary has a mass of $0.095 \pm 0.006$\,\Msun{} \citep{2019_Dupuy, kirkpatrick_2024} and an effective temperature of $T_{\mathrm{eff}} \approx 2100$\,K \citep{2025Huang}. The companion, WISE~J072003.20$-$084651.2~B, is a T5.5 BD with a mass of $66 \pm 4$\,\Mjup, detected at a projected separation of $2.173^{+0.028}_{-0.029}$\,AU, corresponding to an angular separation of approximately $0.32^{\prime\prime}$ \citep{2019_Dupuy}. The system remains unresolved in WISE imaging due to its tight separation.

\begin{figure*}[!htb]
 \centering
\includegraphics[width=0.325\linewidth]{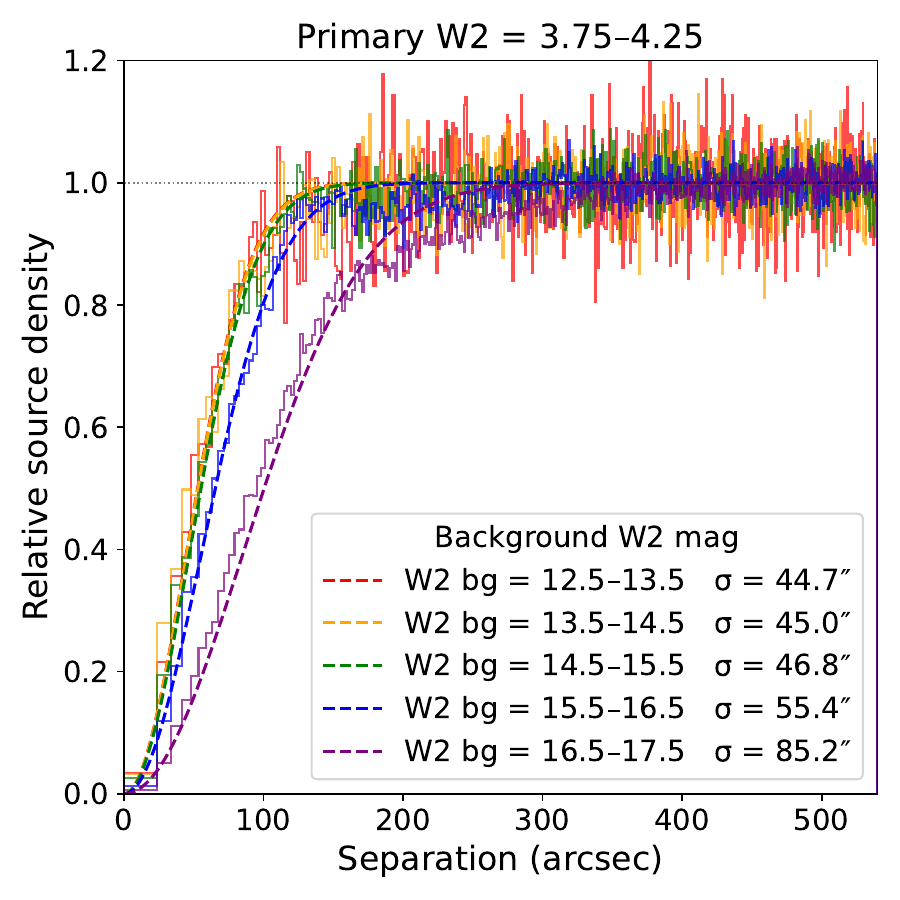}
\includegraphics[width=0.325\linewidth]{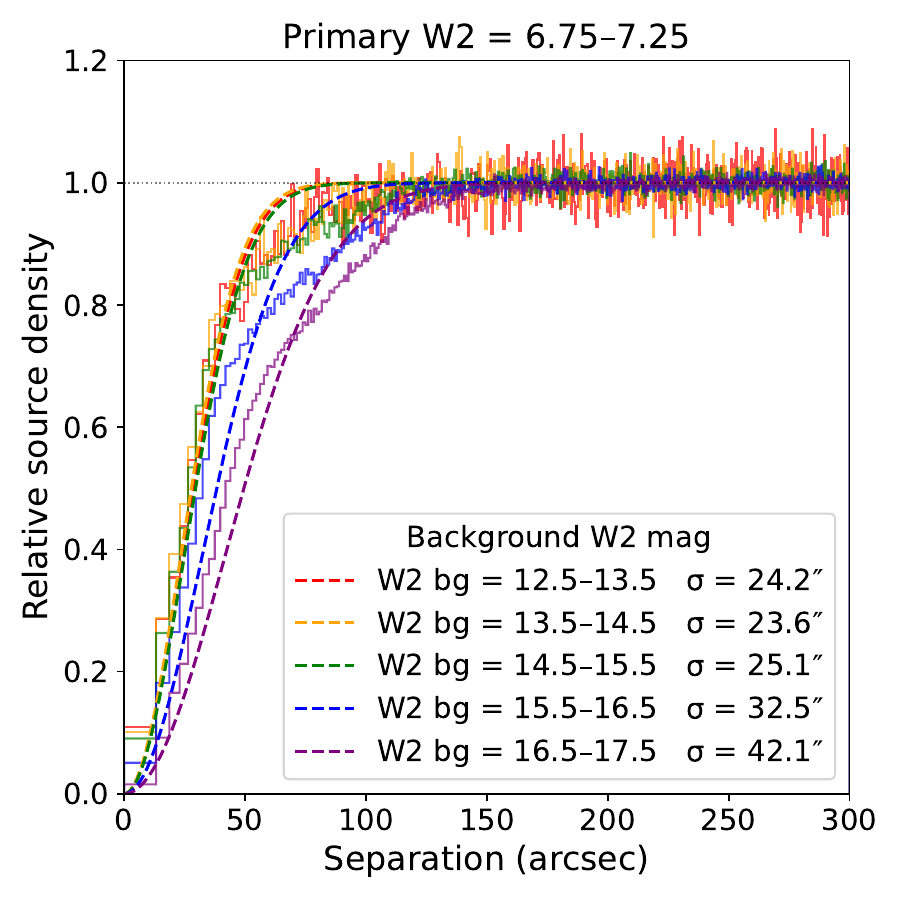}
\includegraphics[width=0.325\linewidth]{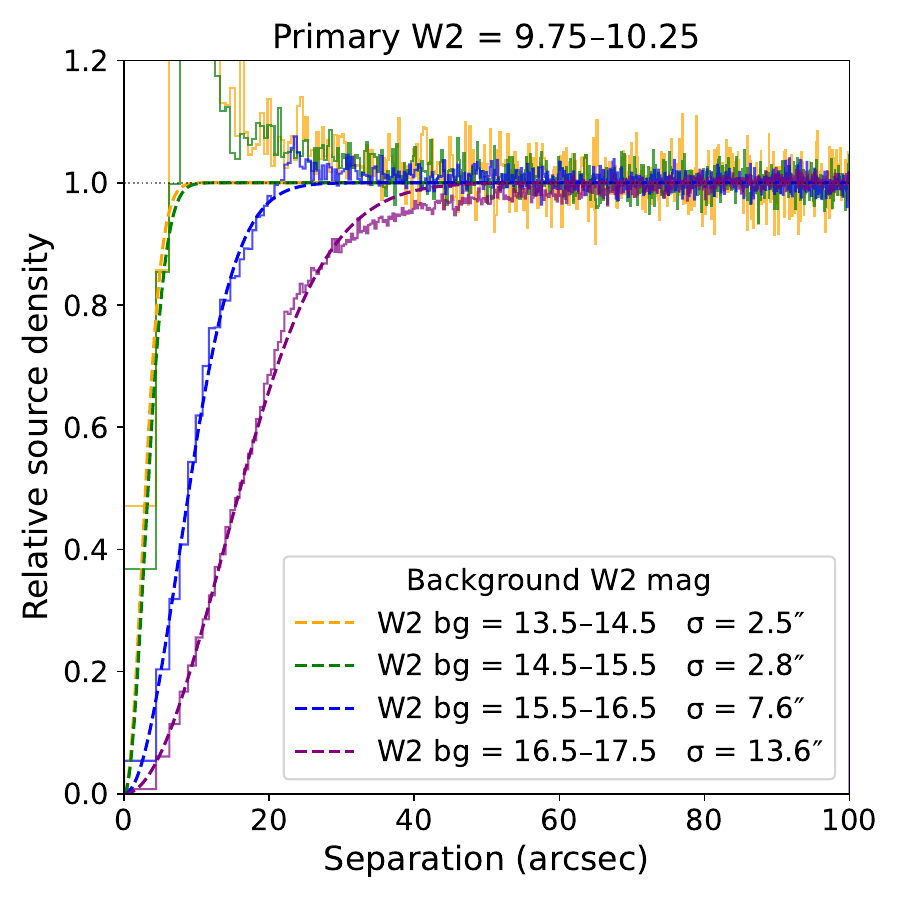}

\caption{Mean background source density as a function of angular distance to $W2=4$, $7$, $10$ stars in CatWISE, corresponding roughly to the 10$^{\rm th}$, 50$^{\rm th}$ and 90$^{\rm th}$ percentile of host stars in our sample. As one would expect, the decline for fainter sources ($W2$=17) happens at larger angular separation to the central object. The decrease is reasonably well parametrized with a Gaussian falloff (dashed lines). In the statistical analysis, values below 0.25 are set to zero as the companion would be embedded in the host star's PSF and challenging to detect without PSF subtraction. We note an increase in visual binaries for brighter companions ($W2=13.5-14.5$ and $14.5-15.5$) around the fainter hosts (left panel). The most plausible explanation being that these are simply physically associated pairs, breaking the assumption of a radially homogeneous distribution of background faint objects.}
\label{fig:sep_w2_cut}
\end{figure*}
\section{Recovery rate at small separation}
The search for co-moving companions is hampered, at small separations, by the modest resolution (FWHM $\sim 6\arcsec$ at $W1$ and $W2$) of WISE and the bright point-spread function of host stars. The falloff in the probability of detecting a companion of a given brightness at a given separation around a host star can be determined empirically. One can reasonably assume that faint sources are intrinsically isotropically distributed around bright stars. Although this is not true on large scales (e.g., galactic plane versus poles), it is a valid 
approximation at the degree scale. 

We compiled the CatWISE catalogue for representative lines of sight uniformly distributed over the entire celestial sphere. For magnitudes ranging from $W2=4$ to $10.5$, in bins of 0.25\,mag, we counted the number of faint sources in annuli of constant area. We then compiled the number of faint background objects at magnitudes of $W2=14$, 15, 16, and 17, representative of substellar companions. {  Figure~\ref{fig:sep_w2_cut} illustrates the falloff of recovery rates close to central stars representative of bright, median, and faint targets in our sample.}

The number of sources is flat at large separations and is normalised to the mean counts at separations of 150 to 300$^{\prime\prime}$. The density of sources falls to half the background value around a separation of $\sim70^{\prime\prime}$ for the $W2=17$\,mag bin. 
The distribution is well fitted with a density $\rho$ modeled as :
\begin{equation}
 \rho(r) = 1-e^{-0.5 r^2/r_0^2} \label{eq5}
\end{equation}

where $r_0$ is the falloff scale and $r$ the angular separation between the central object and the background sources. The $r_0$ value is determined for relevant brightnesses for our stellar sample and companions. The low density of brighter ($W2=10-13$) background objects does not allow for the determination of $r_0$ at these magnitudes; we use the values for $W2=14$ for all brighter companions. This leads to a slightly pessimistic recovery rate for close-in L dwarfs (typically $30-50^{\prime\prime}$). Also, the small number of stars with $W2<4$ does not allow for a direct determination of the recovery rate for the brightest stars in our sample. We use the recovery rate for $W2=4$ for all brighter stars, which tends to slightly overestimate the recovery rate at close-in separations around the few A and F stars in the solar vicinity (65 stars with $W2<3.5$).

\section{Table of recovered companion properties}
\label{appendix:table}

This table compiles all measured parameters for the recovered companions, along with several key properties of the host stars, including their spectral type, the total initial mass of each system, and the parallax.

\begin{table*}
\caption{Parameters of recovered systems with a cool companion}
\label{tab:stellar_parameters}
\centering
\small
\begin{tabular}{lcccccccc}
\toprule
Name & SpType$^{\textcolor{blue}{a}}$ & Mass$^{\textcolor{blue}{b}}$ & Plx$^{\textcolor{blue}{c}}$ & Sep$^{\textcolor{blue}{d}}$ &
Prob$^{\textcolor{blue}{d}}$ &
W$_1$$^{\textcolor{blue}{e}}$ & W$_2$$^{\textcolor{blue}{e}}$ \\
     &                               & [M$_{\text{Jup}}$] & [mas] & [au] & [\%] & [mag] & [mag] \\
\midrule
BD+06 2986 & K8V & 614 & 52.50 $\pm$0.023 & & & & \\
ULAS J150457.65+053800.8 & T6 pec & 41.07$^{\star}$ & & 1287 &  {100} & 16.2 $\pm$0.03 & 14.3 $\pm$0.01 \\
\hline
BD+13 2618 AB & M1.5V & 673 & 86.9 $\pm$0.12 & & & & \\
BD+13 2618 C & T8 & 12.5 $^{+3.2}_{-2.8}$ & & 1175 &   {100}  & 16.37 $\pm$0.03 & 13.89 $\pm$0.01 \\
\hline
BD+21 2486 AB & K7V & 1087 & 50.80 $\pm$0.02$^{\textcolor{blue}{1}}$ & & & \\
2MASS J13054106+2046394 & L5 & 74.45 $\pm$0.24$^{\textcolor{blue}{2}}$ &  & 10075 &  {100}  & 12.56 $\pm$0.01 & 12.191 $\pm$0.008 \\
\hline
G 204-39 & M3V & 415 & 71.49 $\pm$0.01 & & & & \\
2MASS J17580545+4633099 & T6.5 & 28.13 $\pm$7.04$^{\textcolor{blue}{2}}$ & & 2774 &  {100}  & 15.62 $\pm$0.01 & 13.83 $\pm$0.01 \\
\hline
G 48-43 & M3.5V & 239 & 51.002 $\pm$0.02 & & & \\
ULAS J095047.28+011734.3 & T8 & 47 $^{+6.4}_{-7.8}$ & & 1012 &  {100}  & 17.81 $\pm$0.09 & 14.49 $\pm$0.01 \\
\hline
Gl 570 ABC & K4V+M1.5 V & 613 & 169.88 $\pm$0.06 & & & & \\
Gl 570 D & T7 & 36.2 $^{+5.8}_{-5.8}$ &  & 1532 &  {100}  & 14.85 $\pm$0.01 & 12.13 $\pm$0.007 \\
\hline
HD 170573 & K4.5V & 806 & 52.29 $\pm$0.02 & & & &\\
WISE J183207.99-540942.0 & T7 & 35.19$^{\star}$ &  & 11852 &  {99.99}  & 16.57 $\pm$0.08 & 14.437 $\pm$0.03 \\
\hline
HD 65486 & K4V & 775 & 54.15 $\pm$0.01 & & & & \\
2MASS J07580132-2538587 & T4.5 & 28.94 $\pm$8.8$^{\textcolor{blue}{2}}$ & & 1716 &  {100}  & 15.77 $\pm$0.03 & 13.86 $\pm$0.01 \\
\hline
HD 126053 & G1.5V & 1068 & 57.27 $\pm$0.03 & & & \\
ULAS J142320.79+011638.2 & T8 & 39 $^{+7.1}_{-9.1}$ & & 2722 &  {100}  & 17.88 $\pm$0.11 & 14.77 $\pm$0.02 \\
\hline
L 34-26 & M3Ve & 395 & 91.82 $\pm$0.01 & & & & \\
L 34-26 B & T9 & 8$\pm$2$^{\textcolor{blue}{3}}$ & & 6512 &  {100}  & 17.08 $\pm$0.03 & 14.61 $\pm$0.01 \\
\hline
LP 738-14 & M4.5V & 159 & 55.02 $\pm$0.02 & & & &\\
LP 738-14 B & T5.5 & 72$^{+4}_{-7}$$^{\textcolor{blue}{4}}$ &  & 1226 &  {100}  & 16.20 $\pm$0.03 & 14.16 $\pm$0.01 \\
\hline
LP 374-39 & M5V & 156 & 61.65 $\pm$0.06 & & & & \\
2MASS J11225550+2550250 & T6 & 16.58 $\pm$4.56$^{\textcolor{blue}{2}}$ &  & 4131 &  {100}  & 16.26 $\pm$0.042 & 14.10 $\pm$0.02 \\
\hline
Ross 19 & M3.5V & 348 & 57.32 $\pm$0.03 & & & \\
Ross 19 B & T9.5$^{\textcolor{blue}{5}}$ & 15--40$^{\textcolor{blue}{5}}$, 24.10$^\star$ &  & 9902 &  {99.99}  & 18.61 $\pm$0.18 & 15.81 $\pm$0.05 \\
\hline
$\xi$ UMa AB & F8.5V+G2V & 4336 & 114.48 $\pm$0.43$^{\textcolor{blue}{6}}$ & & & & \\
WISE J111838.70+312537.9 & T8.5 & 31.9 $^{+6.4}_{-8.3}$ &  & 4449 &  {100}  & 16.84 $\pm$0.07 & 13.35 $\pm$0.01 \\
\hline
Wolf 940 & M4V & 299 & 80.73 $\pm$0.05 & & & &\\
Wolf 940 B & T8.5 & 28.2$^{+2.6}_{-2.7}$ &  & 389 &  {100}  & 16.51 $\pm$0.03 & 14.37 $\pm$0.019 \\
\hline
Wolf 1130 AB & M1+wd & 8090 & 60.29 $\pm$0.02 & & & &\\
WISE J200520.38+542433.9 & sdT8 & 44.9 &  & 3118 &  {100}  & 17.15 $\pm$0.03 & 15.14 $\pm$0.02 \\
\hline
$\eta$ CrB AB & F9V+G2V & 2204 & 55.98 $\pm$0.78$^{\textcolor{blue}{7}}$ & & & &\\
2MASS J15232263+3014562 & L8 & 44 $\pm$6.46$^{\textcolor{blue}{2}}$ & & 3549 &   {100} & 13.5 $\pm$0.01 & 13.01 $\pm$0.01 \\
\bottomrule
\end{tabular}
\tablefoot{%
The mass of the host star corresponds to the total mass of the system and is also taken from $^{\textcolor{blue}{a}}$~\citet{kirkpatrick_2024}; companion mass is from   
$^{\textcolor{blue}{b}}$~\citet{Zhang2021};  
$^{\textcolor{blue}{c}}$~\citet{gaia_dr3};  
$^{\textcolor{blue}{d}}$~This work;  
$^{\textcolor{blue}{e}}$~\citet{catwise};  
$^{\textcolor{blue}{1}}$~\citet{2021gaia};  
$^{\textcolor{blue}{2}}$~\citet{2023Sanghi};  
$^{\textcolor{blue}{3}}$~\citet{2025Zhang};  
$^{\textcolor{blue}{4}}$~\citet{2012Deacon};  
$^{\textcolor{blue}{5}}$~\citet{2021Schneider};  
$^{\textcolor{blue}{6}}$~\citet{gaia_collaboration_gaia_2018};  
$^{\textcolor{blue}{7}}$~\citet{2007van}.
\\
{
The companion masses (M$_{\text{Jup}}$) were derived following the procedure described in Sect.~\ref{long-term}. The stellar ages adopted to constrain the masses of Ross~19, HD~170573, and BD+06~2986 are 7.49, 2.95, and 3.46~Gyr, respectively. The determination of these parameters follows the analysis presented in Sects.~\ref{const_map} and~\ref{long-term}.}}
\end{table*}

\section{Representative false positives}
\label{appendix:false-positives}
\begin{figure}
    \centering
    \includegraphics[width=0.95\linewidth]{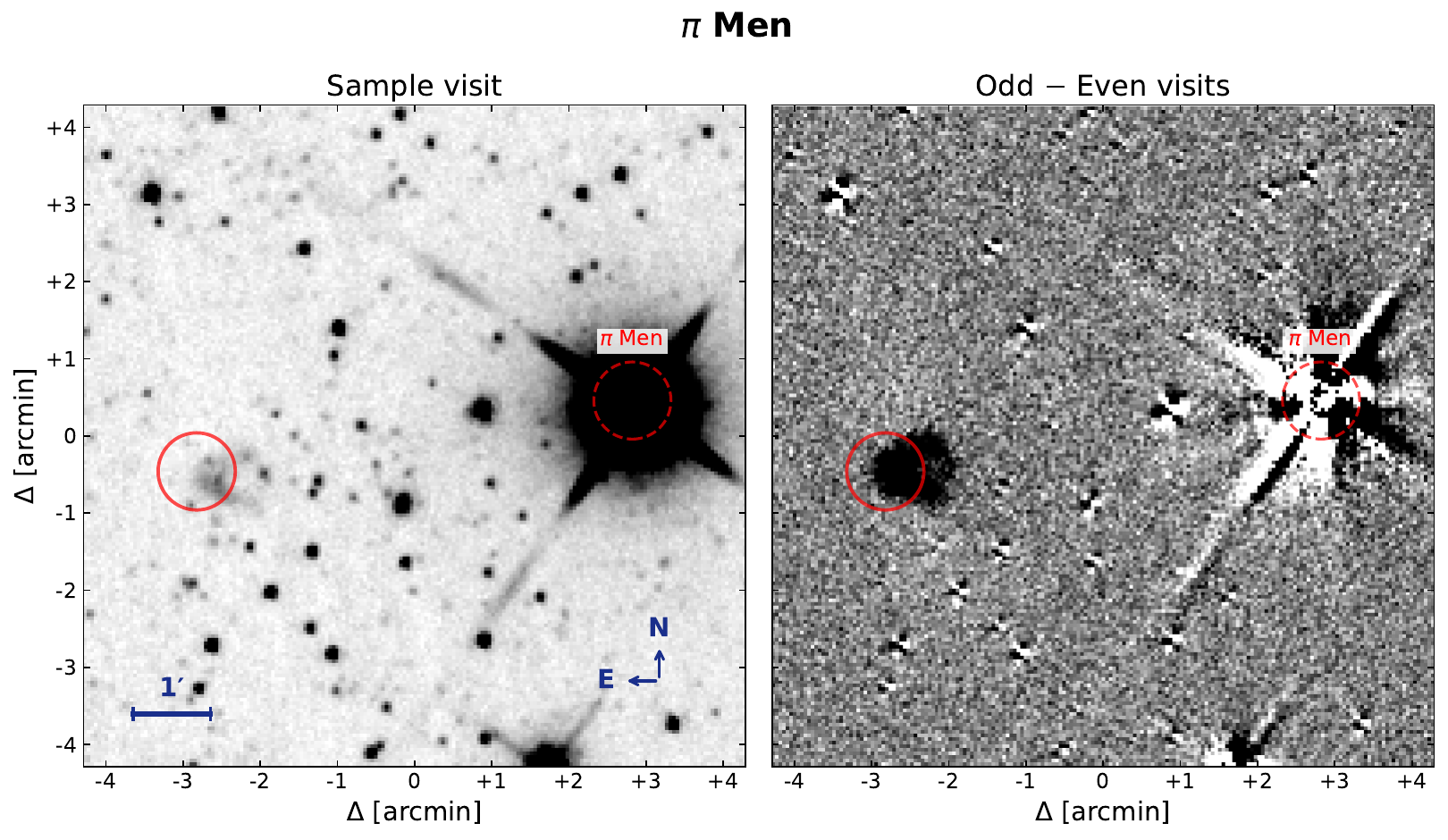}
    \includegraphics[width=0.995\linewidth]{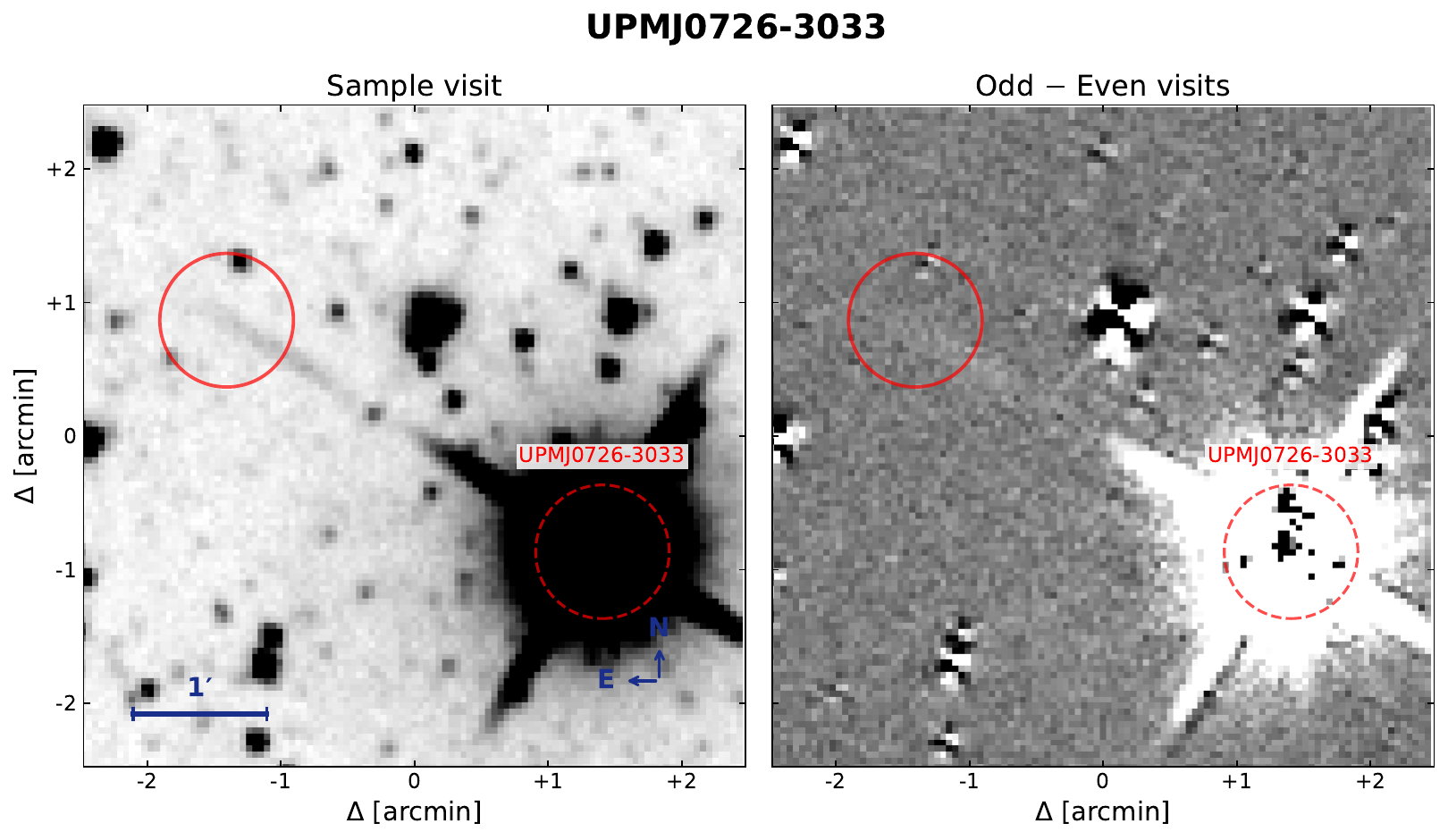}
    \caption{WISE $W2$ cutouts of two false-positive candidates, each shown as a sample individual visit (left) and the difference between median odd' and even' visit frames (right). Red circles mark the host-star position (dashed) and the candidate position (solid). {\it Top:} A candidate around $\pi$\,Men is identified as a ghosting artefact: the odd$-$even frame shows a clear residual at the candidate location, consistent with a ghost pattern that follows the host star and reverses in detector space between visit sets. {\it Bottom:} A candidate around UPM~J0726-3033 is identified as a diffraction spike of the host star, visible as a point-source-like feature in the sample visit at a position consistent with the known spike geometry of the WISE PSF.}
    \label{fig:false-positives}
\end{figure}

All candidates were examined visually in the per-epoch stacks compiled by \citet{Meisner2018b}\footnote{https://unwise.me/}, using the tile index file\footnote{\url{https://portal.nersc.gov/project/cosmo/temp/ameisner/neo8/tr_neo8_index.fits}}. False positives were found be ghosting and diffraction spikes that are co-moving with the star. As the ghosting pattern and diffraction spike properties are wavelength-dependent, some features were present in $W2$ and absent in $W1$ (or present elsewhere in the frame), leading to very red $W2-W1$ colours and co-moving properties (ghosting and spikes `follow' the host star). We examined both the original frames for obvious positional match with diffraction patterns (e.g., a candidate around UPMJ0726-3033 in Figure~\ref{fig:false-positives}). Also, as the WISE spacecraft scanned regions of the sky every six months with a sky-plane orientation flipped by $\sim$180$^\circ$ in detector space. By constructing a median `odd' and `even' visit frame, one can largely subtract the science scene and see if ghost patterns match a tentative candidate (e.g., a candidate around $\pi$\,Men in Figure~\ref{fig:false-positives}).

\section{Projection Effect in the MCMC Framework}
\label{appendix:projection}

The population model describes the true three-dimensional (3D) separation $r$ of binary companions as a power law,
\begin{equation}
    P_{\mathrm{3D}}(r \mid \beta) = \frac{(\beta+1)\, r^{\beta}}{r_{\max}^{\beta+1} - r_{\min}^{\beta+1}}, \qquad r \in [r_{\min},\, r_{\max}],
    \label{eq:pdf_3d}
\end{equation}
while observations only yield the projected (on-sky) separation $\rho$.  For a randomly oriented binary with true separation $r$, the projected separation is $\rho = r\sin\varphi$, where $\cos\varphi$ is uniformly distributed on $[-1,\,1]$.  A change of variables gives the projection kernel
\begin{equation}
    P(\rho \mid r) = \frac{\rho}{r\,\sqrt{r^{2} - \rho^{2}}}, \qquad 0 \le \rho \le r.
    \label{eq:kernel}
\end{equation}
This kernel has an expectation value $\langle\rho\rangle = (\pi/4)\,r \approx 0.785\,r$, peaks near $\rho \approx r$, and has a tail towards small $\rho$ corresponding to systems viewed nearly along the line of sight.

\subsection{Projected separation PDF}

The observed projected-separation probability density function (PDF) is obtained by marginalising the 3D distribution over the projection kernel:
\begin{equation}
    P_{\mathrm{proj}}(\rho \mid \beta) = \int_{\rho}^{r_{\max}} P_{\mathrm{3D}}(r \mid \beta)\;\frac{\rho}{r\,\sqrt{r^{2} - \rho^{2}}}\;\mathrm{d}r,
    \label{eq:pdf_proj}
\end{equation}
where the lower integration limit is $\rho$ because $\rho \le r$ by construction.

\subsection{Discretisation as a matrix--vector product}

Rather than evaluating Eq.~\eqref{eq:pdf_proj} numerically at every MCMC step, we pre-compute a kernel matrix $\mathbf{K}$ of dimension $N_{K}\times N_{K}$ (with $N_{K}=300$) on a logarithmically spaced grid.  We define bin edges $\{r_{j}^{\mathrm{lo}},\,r_{j}^{\mathrm{hi}}\}_{j=1}^{N_{K}}$ for the 3D separation and grid points $\{\rho_{i}\}_{i=1}^{N_{K}}$ for the projected separation, both spanning $[r_{\min},\,r_{\max}]$.  Using the analytic antiderivative
\begin{equation}
    \int \frac{\rho}{r\,\sqrt{r^{2}-\rho^{2}}}\,\mathrm{d}r = -\arcsin\!\left(\frac{\rho}{r}\right) + C,
    \label{eq:antideriv}
\end{equation}
the matrix elements are computed exactly:
\begin{equation}
    K_{ij} =
    \begin{cases}
        \displaystyle \arcsin\!\left(\frac{\rho_{i}}{r_{j}^{\mathrm{lo}}}\right) - \arcsin\!\left(\frac{\rho_{i}}{r_{j}^{\mathrm{hi}}}\right), & \rho_{i} < r_{j}^{\mathrm{lo}}, \\[10pt]
        \displaystyle \frac{\pi}{2} - \arcsin\!\left(\frac{\rho_{i}}{r_{j}^{\mathrm{hi}}}\right), & r_{j}^{\mathrm{lo}} \le \rho_{i} \le r_{j}^{\mathrm{hi}}, \\[10pt]
        0, & \rho_{i} > r_{j}^{\mathrm{hi}}.
    \end{cases}
    \label{eq:kernel_matrix}
\end{equation}
Because $\mathbf{K}$ depends only on geometry, it is independent of $\beta$ and is computed once at startup.  At each MCMC step the projected PDF on the grid is obtained via a matrix--vector product,
\begin{equation}
    \vec{P}_{\mathrm{proj}} = \mathbf{K}\;\vec{P}_{\mathrm{3D}}(\beta),
    \label{eq:matvec}
\end{equation}
where $[\vec{P}_{\mathrm{3D}}]_{j} = P_{\mathrm{3D}}(r_{j}\mid\beta)$.  Values at arbitrary $\rho$ are then obtained by interpolation in $\log\rho$.

\subsection{Role in the likelihood}

The projection enters both terms of the log-likelihood, $\ln\mathcal{L} = \ln\mathcal{L}_{\mathrm{count}} + \ln\mathcal{L}_{\mathrm{pos}}$.

\paragraph{Poisson count term.}
Synthetic companions are drawn with 3D separations $r\sim r^{\beta}$ and projected via $\rho = r\sin\varphi$ (with $\cos\varphi\sim\mathcal{U}(-1,1)$) before evaluating the detection probability $\pi(\rho,m)$.  The expected number of detections is
\begin{equation}
    N_{\mathrm{det}}^{\mathrm{model}} = N_{\star}\,f\,\langle\pi\rangle, \qquad
    \langle\pi\rangle = \frac{1}{N_{\mathrm{MC}}}\sum_{k=1}^{N_{\mathrm{MC}}} \pi\!\left(\rho_{k}^{\mathrm{proj}},\,m_{k}\right),
    \label{eq:Ndet}
\end{equation}
where $N_{\mathrm{MC}}$ is the number of Monte Carlo draws, leading to
\begin{equation}
    \ln\mathcal{L}_{\mathrm{count}} = \ln\,\mathrm{Poisson}\!\left(N_{\mathrm{det}}^{\mathrm{obs}}\;\big|\;N_{\mathrm{det}}^{\mathrm{model}}\right).
    \label{eq:logL_count}
\end{equation}

\paragraph{Position term.}
For each detected companion at observed mass $m_{i}$ and projected separation $\rho_{i}$, the position likelihood is
\begin{equation}
    \ln\mathcal{L}_{\mathrm{pos}} = \sum_{i=1}^{N_{\mathrm{det}}} \ln\!\!\int P(m\mid\alpha)\;P_{\mathrm{proj}}(\rho\mid\beta)\;\pi(m,\rho)\;\mathrm{d}m\,\mathrm{d}\rho
    \;-\;N_{\mathrm{det}}\,\ln\langle\pi\rangle,
    \label{eq:logL_pos}
\end{equation}
where $P_{\mathrm{proj}}(\rho\mid\beta)$ is the projected PDF from Eq.~\eqref{eq:matvec}, \emph{not} the raw 3D power law of Eq.~\eqref{eq:pdf_3d}.  The integral for each companion is evaluated numerically on a small grid centred on its observed $(m_{i},\rho_{i})$, with pre-computed detection probabilities.

\subsection{Summary}

The projection effect is accounted for in two complementary ways:
\begin{enumerate}
    \item Monte Carlo projection ($\rho = r\sin\varphi$) for the count/normalisation term (Eq.~\ref{eq:Ndet}).
    \item Analytic kernel matrix $K$ (Eq.~\ref{eq:kernel_matrix}) for the per-companion position term, reducing each MCMC evaluation to an $\mathcal{O}(N_{K}^{2})$ matrix--vector product plus interpolation.
\end{enumerate}
This approach ensures that the 3D-to-projected mapping is handled consistently throughout the inference, without introducing additional free parameters.

\end{appendix}
\end{document}